\documentclass[sigplan,10pt,screen,nonacm]{acmart}
\renewcommand\footnotetextcopyrightpermission[1]{}
\usepackage[utf8]{inputenc} 
\usepackage[french,english]{babel} % Last language is the default, use 
\usepackage{graphicx} % Pour ins\'{e}rer des images. Utiliser le format jpg pour plus de simplicit\'{e}.
\usepackage{pifont} % Pour utiliser des symboles divers.
\usepackage{xcolor}
\usepackage{comment}
\usepackage{ifthen} % Entrer valeurs bool\'{e}ennes et autres options
\usepackage{amsmath}
\usepackage{amssymb}
\usepackage{amsthm}
\usepackage{siunitx} % \num{1.8e_3} in scientific notation
\usepackage{booktabs}
\usepackage{algorithm}
\usepackage[noend]{algpseudocode}
\usepackage{xspace}
\usepackage{caption}
\usepackage{subcaption}
\usepackage{kbordermatrix} % .sty in root dir

\newcommand{\spores}{\textsc{Spores}\xspace}
\newcommand{\sprinkler}{\textsc{Sprinkler}\xspace}

\newcommand{\por}{\textsc{Por}\xspace}
\newcommand{\pors}{\textsc{Por}s\xspace}
\newcommand{\Trps}{\ensuremath{T_{\text{RPS}}}\xspace}
\newcommand{\vrps}{\ensuremath{\mathcal{V}_{\text{RPS}}}\xspace}
\newcommand{\viewsize}{\ensuremath{l_{\mathcal{V}}}\xspace}
\newcommand{\gossipsize}{\ensuremath{l_{\text{gossip}}}\xspace}
\newcommand{\layer}{\ensuremath{\mathcal{L}}\xspace}

\newcommand{\envelope}{\ensuremath{\mathcal{E}}\xspace}
\newcommand{\cipher}{\ensuremath{\mathcal{C}}\xspace}
\newcommand{\mess}{\ensuremath{\mathcal{M}}\xspace}
\newcommand{\payload}{\ensuremath{\mathcal{P}}\xspace}
\newcommand{\layersize}{\ensuremath{S_{\layer}}\xspace}
\newcommand{\routesize}{\ensuremath{\#_{\layer}}\xspace}
\newcommand{\pubkey}{\ensuremath{pk}\xspace}
\newcommand{\secretkey}{\ensuremath{sk}\xspace}
\newcommand{\Poff}{\ensuremath{P^\text{off}}\xspace}
\newcommand{\picklayer}{\ensuremath{\mathtt{PickLayer}}\xspace}
\newcommand{\timeout}{\ensuremath{T_{\text{out}}}\xspace}

\newcommand{\forwardroute}{\ensuremath{\mathcal{FR}}\xspace}
\newcommand{\backwardroute}{\ensuremath{\mathcal{BR}}\xspace}
\newcommand{\forwardlayer}{\ensuremath{\mathcal{FL}}\xspace}
\newcommand{\backwardlayer}{\ensuremath{\mathcal{BL}}\xspace}
\newcommand{\nspores}{\ensuremath{N_{\textsc{Sp}}}\xspace}
\newcommand{\ntor}{\ensuremath{N_\textsc{Tor}}\xspace}
\newcommand{\nadv}{\ensuremath{N_\text{adv}}\xspace}
\newcommand{\pspores}{\ensuremath{p_{\textsc{Sp}}}\xspace}
\newcommand{\ptor}{\ensuremath{p_\textsc{Tor}}\xspace}
\newcommand{\nusers}{\ensuremath{N_\mathcal{U}}\xspace}
\newcommand{\ndevices}{\ensuremath{N_\text{dev}}\xspace}
\newcommand{\nlocations}{\ensuremath{N_\text{loc}}\xspace}
\newcommand{\nadvusers}{\ensuremath{N^\mathcal{U}_{\text{adv}}}\xspace}
\algnewcommand\algorithmicon{\textbf{on event}}
\algnewcommand\algorithmicrec{\textbf{on receive}}
\algblockdefx[ON]{On}{EndOn}%
  {\algorithmicon\ }%
  {\algorithmicend\ \algorithmicon}
\algblockdefx[RECEIVE]{Rec}{EndRec}%
  {\algorithmicrec\ }%
  {\algorithmicend\ \algorithmicrec}

\makeatletter
\ifthenelse{\equal{\ALG@noend}{t}}%
  {\algtext*{EndOn}\algtext*{EndRec}}%
  {}%
\makeatother

\begin{document}
%%%%%%%%%%%%%%%%%%%%%%%%%%%%%%%%%%%%%%%%%%%%%%%%%%%%%%%%%%%%%%%%%%%%%%%%%%%%%%%%

%don't want date printed
\date{December 2019}

\title[\spores]{\spores: Stateless Predictive Onion Routing for E-Squads}

%for single author (just remove % characters)
\author{Daniel Bosk}
\email{dbosk@kth.se}
\affiliation{%
  \institution{KTH Royal Institute of Technology}
  \city{Stockholm}
  \country{Sweden}
}
\author{Yérom-David Bromberg}
\email{david.bromberg@irisa.fr}
\affiliation{%
  \institution{Univ Rennes, CNRS, Inria, IRISA}
  \city{Rennes}
  \country{France}
}
\author{Sonja Buchegger}
\email{buc@kth.se}
\affiliation{%
  \institution{KTH Royal Institute of Technology}
  \city{Stockholm}
  \country{Sweden}
}
\author{Adrien Luxey}
\email{adrien.luxey@irisa.fr}
\affiliation{%
  \institution{Univ Rennes, CNRS, Inria, IRISA}
  \city{Rennes}
  \country{France}
}
\author{François Taïani}
\email{francois.taiani@irisa.fr}
\affiliation{%
  \institution{Univ Rennes, CNRS, Inria, IRISA}
  \city{Rennes}
  \country{France}
}
% \author{Daniel Bosk
% {\rm Daniel Bosk}\\
% KTH Royal Institute of Technology
% \and
% {\rm Yérom-David Bromberg}\\
% Univ Rennes, CNRS, Inria, IRISA
% \and
% {\rm Sonja Buchegger}\\
% KTH Royal Institute of Technology
% \and
% {\rm Adrien Luxey}\\
% Univ Rennes, CNRS, Inria, IRISA
% \and
% {\rm François Taïani}\\
% Univ Rennes, CNRS, Inria, IRISA
% % copy the following lines to add more authors
% % \and
% % {\rm Name}\\
% %Name Institution
% } % end author

%-------------------------------------------------------------------------------
\begin{abstract}
Mass surveillance of the population by state agencies and corporate parties is now a well-known fact.
Journalists and whistle-blowers still lack means to circumvent global spying for the sake of their investigations.
With \spores, we propose a way for journalists and their sources to plan \emph{a posteriori} file exchanges when they physically meet.
We leverage on the multiplication of personal devices per capita to provide a lightweight, robust and fully anonymous decentralised file transfer protocol between users.
\spores hinges on our novel concept of \emph{e-squads}: one's personal devices, rendered intelligent by gossip communication protocols, can provide private and dependable services to their user.
People's e-squads are federated into a novel onion routing network, able to withstand the inherent unreliability of personal appliances while providing reliable routing.
\spores' performances are competitive, and its privacy properties of the communication outperform state of the art onion routing strategies.
\end{abstract}
%-------------------------------------------------------------------------------

%\settopmatter{printfolios=true} % Required for EUROSYS
\maketitle

%-------------------------------------------------------------------------------

\section{Introduction} % (fold)
\label{sec:introduction}

% \commentAL{Tor is weak in terms of privacy because (1) of the partially centralized 10 directory authorities (DAs) that provide the registry of online relays and (2) because there are only ~6000 relays enabling anonymous routes, whereas there are ~2 million users.
% (1) The 10 DAs must be trusted by the community with the veracity of their registry.
% (2) If an attacker owns the first and last relays of a route, they can de-anonymise the connection; relays are randomly selected; it would be less likely for an attacker to end up in this position if there were more relays $\implies$ every user (and each of their devices) should act as a relay.}
% \commentAL{(2) is not currently in the intro.}

% \commentAL{With Spores (and Tor-SGX), we enable better privacy, because (1) the peer discovery is entirely distributed and (2) our proposal is tailored for very churnful relays i.e. the participants' devices (through fast-paced RPS and \pors).
% We \emph{challenge} Tor's performance (although not killing it), but provide much better privacy properties.
% We need to work on the trades we made, e.g. connectionless routes (no guaranteed order, no integrity check).}

%In this new era of wireless communication, we have come to expect high degrees of synchrony and interactivity.
%People like to exchange pieces of content between one another.

Recent years have been marked by multiple high-profile mass surveillance scandals, involving a diverse range of players, from state agencies~\cite{nakashima_for_2013,macaskill_nsa_2013}, to large technology firms~\cite{cadwalladr_i_2018}, through start-ups with close links to academia~\cite{cadwalladr_i_2018,kosinski_private_2013,youyou_computer-based_2015,matz_psychological_2017}.

In this context, journalists and whistle-blowers must be extremely careful when sourcing or exchanging sensitive or damaging information, but they unfortunately still lack the technical means to fully circumvent corporate and governmental surveillance efforts.
Although data encryption is often used as a first line of defense to protect confidential information, it is unfortunately insufficient on its own to fully protect the parties involved in a remote exchange of data~\cite{granick_2017, HarnikPS10}. Without additional counter-measures, metadata such as a user'
location and activity can usually still be tracked, thus revealing what each user shared with whom and when. 
The exposure of metadata seriously weakens user privacy; as famously emphasized by a former NSA and CIA director: "\emph{We kill people based on metadata.}''~\cite{NaughtonTheGuardian2016}.

Anonymity networks have been proposed to lower the risk of online spying.
They notably hide the identity of the client, most often using mix networks~\cite{Chaum:1981} or onion routing~\cite{goos_hiding_1996,dingledine_tor_2004}.
These approaches basically mangle user requests through a series of relay servers, effectively hiding the link between sender and receiver.
To bootstrap such anonymity networks, relays need to be sufficiently numerous, available, distributed over many autonomous systems, and to provide enough bandwidth.
For instance, Tor~\cite{dingledine_tor_2004}, the most popular onion routing implementation, relies on relays maintained by volunteer operators.
On average, 6000 connected relays handle the traffic of 2 million online users\footnote{See Tor Metrics at \url{https://metrics.torproject.org/}.}.
This small amount of relays relative to the userbase makes Tor particularly prone to attacks such as traffic analysis~\cite{serjantov_passive_2003}.
Tor's infrastructure is also in part centralised, as 10 Directory Authorities (DAs) compute an hourly consensus listing online relays and providing means to contact them.
The DAs constitute a single point of failure, and could be subverted, just like any other relay.
We argue that embracing a peer-to-peer (P2P) strategy---where any participating user also acts as a relay---would notably improve onion routing security by drowning malevolent entities in an ocean of honest peers.

% However, operating such a server is not cheap, such that most implementations tend to have few relays compared to their userbase.
% Tor~\cite{dingledine_tor_2004}, the most popular onion routing implementation, only comprises around 6000 connected relays against 2 million users on average.
% By owning or compromising a sufficient set of servers, and attacker is able to de-anonymise clients through e.g. traffic analysis~\cite{serjantov_passive_2003}.
% Furthermore, Tor has a partially centralised architecture: 10 Directory Authorities (DAs) are trusted by everyone to provide hourly consensus documents containing the list of online relays along with means to contact them.
% These DAs constitute a single point of failure, and are susceptible to attacks just like other relays.
% We believe that building a fully decentralised onion network would improve the users' safety, notably by allowing end-devices to participate to the system as relays.
% By contrast, Tor, one of the most popular anonymous networks, trades lower anonymity guarantees for a much better performance. 
% Tor lower protection arises from a partially centralized design that hinges on a set of \emph{directory servers}, 
% which are used to advertise trusted nodes that are available to bootstrap onion routes. 

We are not the first to advocate more decentralisation, as many attempts at building P2P anonymous data-sharing networks~\cite{Clarke:2001,Freedman:2002,Nambiar:2006,Rennhard:2002,grothoff_gnunet_2017} have been proposed in the last decades.
Typically, such solutions leverage onion routing with added components to perform e.g. node discovery in a decentralised fashion.
Alas, frequent dis/connections of participating peers (\emph{churn}) hinder these systems' performances~\cite{le_blond_towards_2013}, making them unusable in practice.
%Constructing a system that provides fully anonymous content sharing while avoiding the above pitfalls is unfortunately highly challenging. 
% The last two decades did witness many attempts at building P2P anonymous data-sharing networks~\cite{Clarke:2001,Freedman:2002,Nambiar:2006,Rennhard:2002,grothoff_gnunet_2017}.
% These solutions typically leverage onion routing with some improvements to perform node discovery in a distributed manner.
%a step that is otherwise centralized in traditional onion routing schemes. 
%Alas, frequent dis/connections of participating peers (\emph{churn}) hinder these systems' performances~\cite{le_blond_towards_2013}, making them unusable in practice.
% However, because of the high churn typically experienced by P2P networks, 
% the performance of these systems has been discouraging~\cite{le_blond_towards_2013}, 
% as churn causes routes to disappear and be reconstructed frequently, which is particularly costly. 
% By contrast, Tor, one of the most popular anonymous networks, trades lower anonymity guarantees for a much better performance. 
% Tor lower protection arises from a partially centralized design that hinges on a set of \emph{directory servers}, 
% which are used to advertise trusted nodes that are available to bootstrap onion routes. 

In this paper, we propose \spores, a fully decentralised anonymous file exchange protocol, adapted from traditional onion routing.
To sustain the unavoidable churn, we revisit P2P by leveraging machine learning in order to predict peers' availability. 
Towards this goal, we make two assumptions on the peers: 
we consider that each participating device belongs to a particular user, 
and that each user owns several devices (encouraged by the multiplication of appliances per household~\cite{noauthor_cisco_2019}).
We thus propose the concept of \emph{e-squads}: 
an e-squad is constituted of a single user's devices, that model their user's behaviour by exchanging information through gossip messaging.
Using their user behavioural model, each device can predict its future state of availability.
%By publishing their future availability estimates globally, devices enable the whole system to predict failures and better resist churn.

We use these estimates to propose a novel onion routing mechanism, called \emph{Probabilistic Onion Routing} (\por).
%Using these estimates, we propose an upgrade of traditional onion routing, called Probabilistic Onion Routing (\por).
With \por, onion routes may include several candidate relays at each hop, such that a message can go through the route as long as one candidate is online per hop.
Users employ the availability prediction to ensure that the routes they create will remain available with a good probability, without sacrificing their privacy.
In addition, \por is stateless---all routing information is contained in the headers---enabling short-lived relay servers to pass on messages as soon as they join the network, eschewing any bootstrap phase.

%In the end,
Building on \por, \spores enables two users to exchange a file in complete anonymity:
firstly, the two users exchange file metadata and routing information \emph{out-of-band} (using another communication channel than \spores);
then, using the routes they agreed upon, their respective e-squads collaborate to perform the file exchange through \spores, without revealing their identities to the rest of the network.
The proposed service is quite similar to OnionShare\footnote{An anonymous file exchange service backed by Tor's hidden services, see \url{https://onionshare.org/}.}, 
without the security limitations of Tor, and without the need to spawn a web service prior to the exchange.
We say that \spores is an \emph{a posteriori} file exchange service.

Our contributions are the following:
\begin{itemize}
	\item We introduce the concept of \emph{e-squads}, and build a intra-e-squad protocol, that allows devices owned by the same user to create user behavioural models and thus, to %% compute and publish an
    estimate their future availability. 
	\item Based on the e-squad predictions, we introduce Probabilistic Onion Routing (\por), a onion routing protocol %% and message format
    tailored for networks with high degrees of churn.
	\item We use \por to realise \spores, an anonymous file transfer service. After an initial out-of-band exchange of metadata, two users can privately exchange a file. %% between their e-squads
  The transfer remains efficient despite the network's unreliability, and ensures a better anonymity than existing onion routing approaches.
% 	\item We propose a novel message format for set-based onion routing which allows each layer of an onion route to contain several alternative next nodes.
% 	\item Building upon it, we introduce Probabilistic Onion Routing (\por), a protocol that allows stateless onion routing of messages.
% 	The several candidate relays per hop, along with its statelessness, make \por tailored for networks with high churn: 
% 	a route remains available as long as a single node is online per layer, and nodes need no bootstrap phase to start forwarding messages.
% % 	\item Building on it, we introduce a Probabilistic Onion Routing (\por) protocol that, 
% % given a secure random-peer sampling (RPS) protocol and availability estimates, 
% % selects nodes to populate the layers of the header in order to maximize the availability of the route. 
% % This node selection procedure yields reliable routes even in the presence of low-availability nodes while retaining the strong privacy properties of onion
% % routing.
% 	\item We provide a protocol for an e-squad to predict the availability of its devices, using Markov models and e-squad overlays (based on \sprinkler). 
% 	\item We use \por to realize \spores, a privacy-preserving file transfer service among users. 
% 	\spores uses the e-squads' availability predictions to craft reliable probabilistic onion routes despite the low-availability relays, 
% 	and eschews any third-party storage or centralized bootstrapping service.
\end{itemize}

The remainder of this chapter is organised as follows: 
we first present our protocol and its sub-systems in section~\ref{sec:spores_approach}, before presenting our attack model and security properties in section~\ref{sec:attack_model}.
An evaluation of \spores is proposed in section~\ref{sec:evaluation}.
We %% discuss the properties and limitation of our proposal in section~\ref{sec:security_discussion},
make a review of the state of the art in section~\ref{sec:related_works}, and finally conclude the paper in section~\ref{sec:conclusion}.
\section{Our approach} % (fold)
\label{sec:spores_approach}

\begin{figure}[ht]
\centering
\def\svgwidth{0.75\linewidth}
%% Accompanies image file 'outline.pdf' (pdf, eps, ps)
\begingroup%
  \makeatletter%
  \providecommand\color[2][]{%
    \errmessage{(Inkscape) Color is used for the text in Inkscape, but the package 'color.sty' is not loaded}%
    \renewcommand\color[2][]{}%
  }%
  \providecommand\transparent[1]{%
    \errmessage{(Inkscape) Transparency is used (non-zero) for the text in Inkscape, but the package 'transparent.sty' is not loaded}%
    \renewcommand\transparent[1]{}%
  }%
  \providecommand\rotatebox[2]{#2}%
  \newcommand*\fsize{\dimexpr\f@size pt\relax}%
  \newcommand*\lineheight[1]{\fontsize{\fsize}{#1\fsize}\selectfont}%
  \ifx\svgwidth\undefined%
    \setlength{\unitlength}{345.40064053bp}%
    \ifx\svgscale\undefined%
      \relax%
    \else%
      \setlength{\unitlength}{\unitlength * \real{\svgscale}}%
    \fi%
  \else%
    \setlength{\unitlength}{\svgwidth}%
  \fi%
  \global\let\svgwidth\undefined%
  \global\let\svgscale\undefined%
  \makeatother%
  \begin{picture}(1,0.85869591)%
    \lineheight{1}%
    \setlength\tabcolsep{0pt}%
    \put(0,0){\includegraphics[width=\unitlength,page=1]{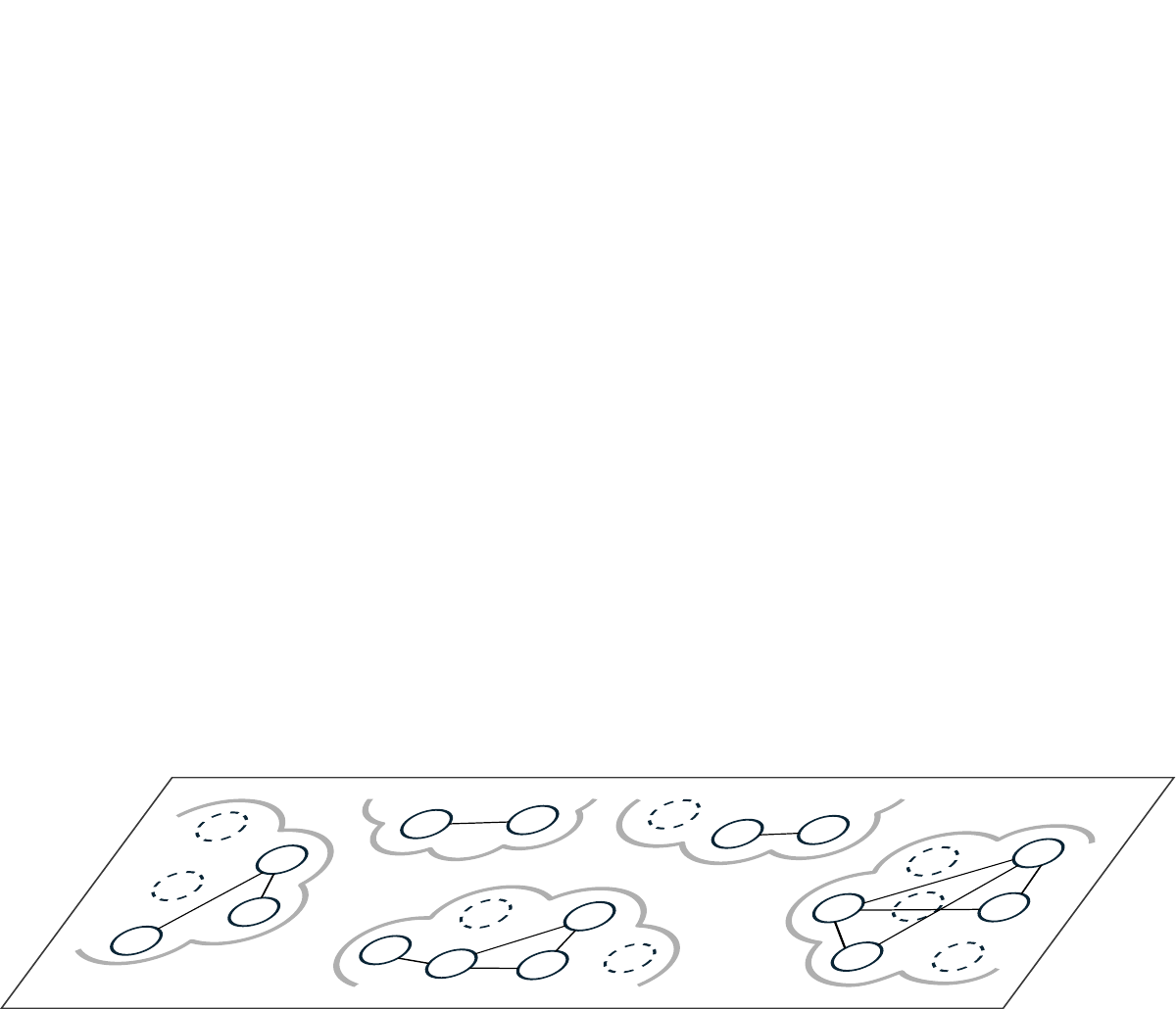}}%
    \put(-0.00226983,0.16707203){\color[rgb]{0,0,0}\makebox(0,0)[lt]{\lineheight{0}\smash{\begin{tabular}[t]{l}\ref{sub:the_e_squad_overlay}\end{tabular}}}}%
    \put(0,0){\includegraphics[width=\unitlength,page=2]{figures/outline.pdf}}%
    \put(-0.00226983,0.38745594){\color[rgb]{0,0,0}\makebox(0,0)[lt]{\lineheight{0}\smash{\begin{tabular}[t]{l}\ref{sub:the_global_overlay}\end{tabular}}}}%
    \put(0,0){\includegraphics[width=\unitlength,page=3]{figures/outline.pdf}}%
    \put(-0.00226982,0.60787619){\color[rgb]{0,0,0}\makebox(0,0)[lt]{\lineheight{0}\smash{\begin{tabular}[t]{l}\ref{sub:spor_stateless_probabilistic_onion_routes}\end{tabular}}}}%
    \put(0,0){\includegraphics[width=\unitlength,page=4]{figures/outline.pdf}}%
    \put(-0.00226983,0.82833275){\color[rgb]{0,0,0}\makebox(0,0)[lt]{\lineheight{0}\smash{\begin{tabular}[t]{l}\ref{sub:spores_file_exchanges_through_spor}\end{tabular}}}}%
  \end{picture}%
\endgroup%

\caption{The subsystems constituting \spores.
From bottom to top: the e-squad overlays (Sec.~\ref{sub:the_e_squad_overlay}) and the global overlay (\ref{sub:the_global_overlay}) enable Probabilistic Onion Routing (\ref{sub:spor_stateless_probabilistic_onion_routes}), which lies at the core of the \spores anonymous file exchange service (\ref{sub:spores_file_exchanges_through_spor}).}
% the e-squad overlay enabling communication inside each e-squad, presented in Sec.~\ref{sub:the_e_squad_overlay}; 
% the global overlay allowing all devices to reach each other, described in Sec.~\ref{sub:the_global_overlay};
% the Stateless Probabilistic Onion Routes (\pors) protocol, that will be covered in  Sec.~\ref{sub:spor_stateless_probabilistic_onion_routes};
% finally, we will present \spores, the anonymous e-squad file exchange protocol in Sec.~\ref{sub:spores_file_exchanges_through_spor}.}
\label{fig:outline}
\end{figure}

\spores is constituted of several sub-systems, as depicted in Fig.~\ref{fig:outline}.
At the root of our proposal is predictive routing, which is made possible by having every peer publish predictions about their future connectivity.
Each peer in the network is a device owned by an individual, who possesses several appliances.
Their appliances exchange information about their user through a \emph{private e-squad overlay} (Sec.~\ref{sub:the_e_squad_overlay}); there is one private overlay per participating user.
Devices use this information to build a model of their user, and make predictions about their future availability.
This estimate is regularly published by each device, along with their network address and public key, on the \emph{global overlay} (Sec.~\ref{sub:the_global_overlay}).
This overlay enables peer discovery at the scale of the whole network.
Using the above information, any device can intelligently build Probabilistic Onion Routes (\pors, see Sec.~\ref{sub:spor_stateless_probabilistic_onion_routes}), a new kind of onion route featuring several relays per hop, in order to maximise the route's availability despite the relays' churn.
Finally, Sec.~\ref{sub:spores_file_exchanges_through_spor} presents the anonymous file exchange protocol in itself, \spores, that anonymises a file transfer through \pors, while making use of each user's e-squad for increased dependability.

% Each user owns several devices that participate in the network, that we call the user's \emph{e-squad}.

% %We now detail the operation of our anonymous file-sharing service for e-squads.
% As shown on Fig.~\ref{fig:outline}, from bottom to top, \spores is constituted of two network overlays, the Probabilistic Onion Routing (\por) protocol, and the file exchange service.
% The latter creates probailistic onion routes (\pors) to exchange files in a private way.
% The devices' addresses constituting the onion routes are collected using the global overlay.
% Routes are only created \emph{once}, during the file exchange bootstrap.
% The e-squad overlays compute each device's probability of remaining online in the near future, which allows the route creation process to try to maximize the future availability of the routes.

% We present each sub-system from the ground up: we first introduce the e-squad overlay and its several purposes, before covering the global overlay.
% We then specify the \por protocol, before finally diving into our anonymous file exchange service.

%the e-squad's and the global one, that are presented in Sec.~\ref{sub:the_e_squad_overlay} and \ref{sub:the_global_overlay} respectively.
%Our onion routing protocol, labelled \por, is presented in Sec.~\ref{sub:spor_stateless_probabilistic_onion_routes}.
%Enabled by the overlays and the routing protocol, we present the file exchange system, \spores, in Sec.~\ref{sub:spores_file_exchanges_through_spor}.

\subsection{The private e-squad overlay} % (fold)
\label{sub:the_e_squad_overlay}

An e-squad overlay is constituted only of devices owned by the same user.
Its role is to make any user-related information available to the whole set of devices.
%We first describe the gossip protocol, before proposing the behavioural model computed by each device.

% At the root of \spores is predictive routing.
% Devices must predict and publish their own probability of remaining online, which depends on their user's behaviour.
% This information is then used to create dependable stateless onion routes (\pors).

% Devices thus need to collect information on their user's behaviour to make such predictions.
% To this end, we employ a gossip algorithm among the e-squad, to enable all devices to aggregate their user data over time.

% % As was seen in the previous chapters, devices need to collect information on their user to make such predictions.
% % For that reason, we build yet another variation of the \sprinkler Gossiper protocol presented in section~\commentAL{Ref to manuscript chapter}, that lets each e-squad device learn the global sequence $S$ of their user's behaviour by gossiping every new update among them.

% In the e-squad overlay, devices inform each other every time they turn on (which will allow them to model their user's behaviour), 
% but also share every file exchange information their user participates in (as an uploader or receiver).
% Although file exchanges only have one sender and one receiver device, the latter information allows any e-squad member to receive chunks and acknowledgements while the end-device is offline (more on the topic in part~\ref{sub:spores_file_exchanges_through_spor}).

\subsubsection{Sharing the user's behaviour} % (fold)
\label{ssub:sharing_the_user_s_behaviour}

The e-squad overlay is based on the \sprinkler Gossiper algorithm~\cite{luxey_sprinkler_2018}, extended with acknowledgements~\cite{bromberg_cascade_2018} to better resist churn.
As in \sprinkler, we assume each user owns a set of devices $\mathcal{D}$. The user's activity is an ever-growing sequence $S = \left\{r_1, \dots, r_i, \dots\right\}$ of interactions $r_i$.
Each device $d$ initially only knows about interactions that took place on it, $S_d$.
Through the e-squad overlay, all nodes of the e-squad share their local interactions to obtain $S = \bigcup_{d \in \mathcal{D}} S_d$.
Interactions are timestamped, and totally ordered. %no two interactions happen at once, such that there is a total order of the interactions in the sequence.

Contrarily to \sprinkler, an interaction can be either a device usage event or a file exchange event.
A file exchange is tied to a single device (sender or receiver), and all the e-squad needs to know what device is involved in which file exchange.
To this end, an interaction $r$ is constituted of the following fields:
$$r = (ts, d, \mathtt{typ}, f) \in \mathbb{R} \times \mathcal{D} \times \mathcal{T} \times \mathcal{F} $$
such that: $ts \in \mathbb{R}$ is the interaction timestamp, $d \in \mathcal{D}$ is the descriptor (see Sec.~\ref{sub:the_global_overlay}) for the device where the interaction $r$ took place, $\mathcal{T} = \left\{\text{USE}, \text{DL}, \text{UL}\right\}$ is the set of interaction \emph{types} (resp. device usage, new file download, or new file download). 
When \texttt{typ} = DL (resp. UL), $f \in \mathcal{F}$ contains the unique ID of the file that just started downloading (resp. uploading) on $d$.
When \texttt{typ} = USE, it means that device $d$ was connected at time $ts$.
Devices issue a USE message when they are grabbed, and every $T$ seconds while they remain connected.

\subsubsection{Modelling the user's behaviour} % (fold)
\label{ssub:modelling_the_user_s_behaviour}

% Each device $d$ builds a \emph{discrete-time} Markov chain of the user's behaviour given the sequence of usage $S^U = \left\{ r \in S, r.\mathtt{typ} = \text{USE}\right\}$.
% They use it to compute their own probability $P_d$ of staying online in the near future, that they will publish to the overall network.

Given the sequence of devices' usage $S^U = \left\{ r \in S, r.\mathtt{typ} = \text{USE}\right\}$, 
each device needs to compute its own probability $P_i(d)$ of staying online in the near future, before advertising it.%that they will publish to the overall network.
% We propose a simplistic model, that could be much improved using more user data and a more complex prediction.
% Still, we demonstrate in our evaluation that a simple prediction like the following suffices at providing convincing improvements for the \pors availability.

% \begin{table}[t]
% \caption[caption]{Sample of an e-squad availability sequence $X$.%% , as a 2D boolean matrix.
%   \label{tab:connection_times}}
% \vspace{3pt}
% \centering
% \begin{tabular}{@{}lccccc@{}} \toprule
% 		& $\cdots$	& $X_{i-1}$	& $X_i$		& $X_{i+1}$	& $\cdots$ \\ \midrule
% Laptop	& $\cdots$	& 1			& 0			& 0			& $\cdots$ \\
% Phone	& $\cdots$	& 1			& 1			& 1			& $\cdots$ \\
% TV		& $\cdots$	& 0			& 1			& 0			& $\cdots$ \\ \bottomrule
% \end{tabular}
% \end{table}

First of all, using only $S^U$, each device builds an availability sequence $X = X_1, \dots, X_i, \dots$,
where $X_i$ contains the set of online devices during the interval $\left[t_i, t_{i+1} \right[$ 
(see Eq.~\ref{eq:avail_sequence}).
The observation sequence has a period of $T$: $\forall i, t_{i+1} = t_i + T$.
The sequence $X$ can be represented as a 2D sparse matrix of booleans.
%as shown in the table~\ref{tab:connection_times}, 
%that represents the devices usage of a user owning a laptop, a phone, and a smart-TV.
\begin{equation}	
X_i(d) = 1 \iff \exists r \in S^U, r.d = d \wedge t_i \leq r.ts < t_{i+1}.
\label{eq:avail_sequence}
\end{equation}
Now, to predict $P_i(d)$, we consider that the stochastic process $X$ follows the Markov property: `the future only depends on the present, not on the past'.
We use the hypothesis in Eq.~\ref{eq:p_d_markov}.

As a result, the probability for $d$ to be online in the near future only depends on its probability to stay online after the current round $X_i = x$.
To estimate this probability, we simply count\footnotemark the number of times the current situation $x$ led to a situation where $d$ was also online (Eq.~\ref{eq:p_d_addition}):
\footnotetext{%
Because we work with low-probability events observed with small amounts of data, there is a possibility that an event never occurs in $X$. 
To counter that, we apply \emph{add-one smoothing}~\cite{russell_artificial_2003} while computing probabilities. 
We left this engineering optimization out of the demonstration for clarity.}
\begin{align}
P_i(d)	& = P\left[ X_{i+1}(d)=1 \mid X_i = x, \dots, X_0 = x_0\right] \nonumber \\ 
		& = P\left[ X_{i+1}(d)=1 \mid X_i = x\right] \label{eq:p_d_markov}\\
		& = \frac{ \left|\left\{ X_j \in X, X_j = x \wedge X_{j+1}(d) = 1 \right\}_{0 \leq j < i}\right| }%
				 { \left|\left\{ X_j \in X, X_j = x \right\}_{0 \leq j < i}\right| } \label{eq:p_d_addition}
	%& = \sum\limits_{\substack{j < i \\ X_j = x \wedge X_{j+1}(d) = 1}} P\left[ S_{i+1}=x' \mid S_i=x \right]
\end{align}

Given the high dimensionality of the state space, it might happen that $x$ was never seen before, leading to an undefined $P_i(d)$
In such a case, we estimate the probability that $d$ stays online two turns in a row as fallback.

% Alas, given the high dimensionality of the state space, it is very likely that $x$ was never seen in $X$ until now, leading to an undefined $P_i(d)$.
% In such a case, we fall back on $P'_i(d)$, an estimate of the probability that $d$ stays online two turns in a row:
% \begin{equation}
% P'_i(d) = \frac{ \left|\left\{ X_j \in X, X_j(d) = 1 \wedge X_{j+1}(d) = 1 \right\}_{0 \leq j < i}\right| }%
% 		    { \left|\left\{ X_j \in X, X_j(d) = 1 \right\}_{0 \leq j < i}\right| } \label{eq:p_d_prime}
% \end{equation}
% When $x$ was never observed, $P_i(d)$ takes the value of $P'_i(d)$. 

% subsubsection modelling_the_user_s_behaviour (end)

% subsection the_e_squad_overlay (end)

\subsection{The global overlay} % (fold)
\label{sub:the_global_overlay}

To creates \pors, each device needs to know some other online devices' descriptors.
For a device $d$, a descriptor contains its address $@_d$, its public key $\pubkey_d$, and its estimated probability of remaining online $P_i(d)$.
$d$ also knows its own private key $\secretkey_d$, that it uses to decipher messages encrypted with $\pubkey_d$.

%To create \pors, devices need to know the address, encryption key, and probability of remaining available of some nodes in the system.
Given the decentralised nature of \spores, we cannot rely on a central registry of online peers as e.g. Tor does.
We use instead a global Random Peer Sampling (RPS) service~\cite{jelasity2007gossip,voulgaris_cyclon:_2005}.
Essentially, each node maintains a view \vrps containing \viewsize other devices’ descriptors. 
Every \Trps seconds, the view is updated as follows: 
a device $d$ pops the oldest descriptor $d'$ from its view, 
then swaps a predefined number of \gossipsize elements from \vrps with $d'$. 
Both devices add a fresh descriptor of themselves to the view exchange. 
If $d'$ was offline, its descriptor is simply removed from $d$’s view, with no further modification to \vrps.

This allows for two things: 
firstly, each device’s view contains a constantly changing random sample of participating devices; 
secondly, stale descriptors get removed from one's view after a bounded time, such that \vrps mostly contains online devices' descriptors.

Given their epidemic nature, RPS services are very sensitive to Byzantine attacks, 
where malicious nodes gossip bad views in order to disrupt the randomness of the neighbourhood graph.
Several proposals overcome this limitation, sometimes by relying on a trusted third-party~\cite{bakker_puppetcast_2008},
sometimes by computing a reputation of the peers~\cite{bortnikov_brahms_2009,jesi_secure_2010}.
%validity of received messages~\cite{bortnikov_brahms_2009,jesi_secure_2010}.
We leverage on the latter, so as to remain entirely decentralised.

% In \spores, each device $d$ computes its own asymmetric key pair $(\pubkey_d, \secretkey_d)$, and publishes its descriptor to remote peers. It contains:

% \begin{itemize}
% 	\item the node's address $@_d$;
% 	\item its public key $\pubkey_d$;
% 	\item its estimated probability of staying online $P_i(d)$.
% \end{itemize}

% subsection the_global_overlay (end)

\subsection{\pors: Probabilistic Onion Routes} % (fold)
\label{sub:spor_stateless_probabilistic_onion_routes}

\paragraph{Legacy Tor primer} Onion routing makes connections between a client (say Alice) and their correspondent (Bob) go through two or more servers (or relays) before reaching their destination.
With Tor, to create a route, Alice randomly picks three relays to constitute the path, and incrementally establishes TLS connections to each of them through the route.
Once the route is established, it constitutes a persistent two-way TCP stream, although the traffic is internally chunked into fixed-size messages (or cells).
Cells contain a header and a payload, that are encrypted altogether by the client several times: once per relay.
Upon reception of a cell from the sender to its destination, each relay deciphers it using the encryption keys negotiated during the TLS connection bootstrap.
Bob finally receives the message originally written by Alice, and can answer back on the same pipe.
Messages on this direction are incrementally encrypted by the relays, such that Alice receives Bob's message hidden under three layers of encryption.
She decrypts it using the keys that were negotiated with the relays during the connection establishment.

% The traffic is cut into fixed-size messages (or cells)
% The traffic is cut in messages, is encrypted several times by the source, such that each node (or \emph{relay}) on the path has to decipher a \emph{layer} of encryption---hence the onion analogy---revealing only the address of the next hop and the encrypted remainder of the header, to be forwarded.
% At the end, the recipient receives the payload, that only they can decrypt.
% We use the terms hop and layer interchangeably, even though they respectively represent the network and header structure, as they serve the same purpose.

The anonymizing property stems from the fact that each hop $\layer_i$ only knows the address of the previous relay $\layer_{i-1}$ (that sent the message) and the address of the next $\layer_{i+1}$ (determined at the connection's establishment).
Given that routes contain two or more hops, no intermediary knows both the sender and the receiver of a message, thus making the communication anonymous.
%In practice, Tor uses three hops, notably for statistical security (one has to own the three intermediaries to easily de-anonymize the message).

\begin{figure}[t]
	\centering
	\includegraphics[width=0.8\columnwidth]{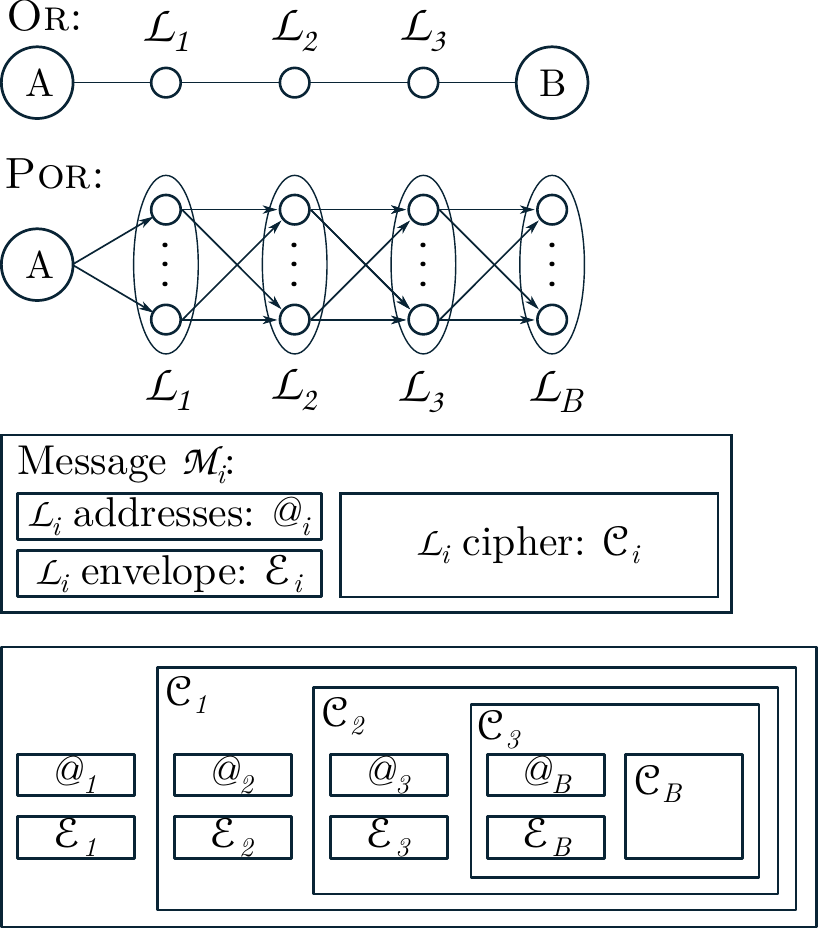}
	\caption{In Onion Routing (OR), each layer $\layer$ is constituted of only one node.
	In Probabilistic Onion Routing (\textsc{Por}), there are several \emph{candidate} nodes that each message can go through at each layer.
	The bottom part of the plot shows the format of a \por message, and the message effectively sent by Alice to relays in $\layer_1$.}
	\label{fig:onion_routes_and_message}
\end{figure}

% \begin{figure*}[t]
% \centering
% \begin{minipage}[t]{.49\linewidth}
%   \centering
%   \includegraphics[width=.7\linewidth]{figures/or_vs_por.pdf}
%   \caption{Two different onion ways for Alice ($A$) to anonymously communicate with Bob ($B$). In Onion Routing (\textsc{Or}), each layer $\layer$ is constituted of only one node.
% 	In Probabilistic Onion Routing (\textsc{Por}), there are several \emph{candidate} nodes that each message can go through at each layer.\label{fig:or_vs_por}}
% \end{minipage}\hfill%
% \begin{minipage}[t]{.49\linewidth}
%   \centering
%   \includegraphics[width=.8\linewidth]{figures/por_message.pdf}
%   \caption{Format of a \por header (on top), and the full header as sent by Alice in Fig.~\ref{fig:onion_routes_and_message}.
%   Each cipher $\cipher_i$ is encrypted using a symmetric key hidden in the envelope $\envelope_i$, only readable by members of layer $\layer_i$.
%   $\cipher_B$ contains the message payload.
%   \label{fig:por_message}}
% \end{minipage}
% \end{figure*}

\paragraph{Several relays per layer} 
The basic idea of Probabilistic Onion Routing (\por) is depicted in Fig.~\ref{fig:onion_routes_and_message}:
each message sent from Alice to Bob may pass through multiple candidate nodes at each hop, instead of only one in traditional onion routing.

In onion routing, when any of the relays becomes unavailable, the route is broken and a new one needs to be created.
The prime interest of \pors is that they are resilient to intermediaries churn:
we only need one online relay per layer for the route to function.
In practice, when a node from layer $\layer_i$ has a message to transmit, it tries sending it to each device in layer $\layer_{i+1}$ in random order, until it succeeds or all attempts fail.
In the latter case, the message is dropped.
%Between each attempt, the node waits a timeout before considering a node offline.
%Only if all nodes from $\layer_{i+1}$ are offline, is the message dropped.
%  one from layer $\layer_{i+1}$, tries to send it the message, and node from picks nodes from layer $\layer_{i+1}$ until it successfully forwards the message.
% The latter is dropped only when none of $\layer_{i+1}$'s relays are available.

In contrast to Tor, \pors do not create TLS connections, which would be inapplicable with several nodes per hop.
Instead, routes are stateless: all routing information is contained inside an encrypted message.

%This calls for new \emph{group} encryption primitives that we discuss below.

\paragraph{\por messages}
Fig.~\ref{fig:onion_routes_and_message} shows the format of \por headers (along with an example of a full message as sent by Alice in Fig.~\ref{fig:onion_routes_and_message}).
A \por message $\mess_i$, as received by any member of $\layer_i$, is constituted of three parts, $\mess_i = \left(@_{i}, \envelope_{i}, \cipher_{i}\right)$:

\begin{itemize}
	\item $@_{i}$: The addresses of all members of the current layer $\layer_i$, used by nodes of $\layer_{i-1}$ to forward $\mess_i$. %the present message;
	\item $\envelope_{i}$: An envelope, destined to $\layer_i$, that will allow them to decrypt the cipher $\cipher_{i}$.
	%, as will be detailed in the next paragraph.
	\item $\cipher_{i}$: A cipher, that can be deciphered by any member of $\layer_i$ using $\envelope_{i}$. 
	It can unravel into another \por message $\mess_{i+1}$ for the next layer $\layer_{i+1}$, or into an application payload once the message reached its destination.
\end{itemize}

By getting rid of TLS connections in favour of header-based routes, \por enables \emph{stateless} routing:
no prior communication is needed with relays to establish onion routes, they simply decipher any received message, and read their header to forward them to the next layer.
This is particularly interesting for short-lived nodes such as seldom connected personal devices as we target: 
they can participate in the system as soon as they join, without any bootstrap phase.
Their disconnection does not mandate a new route construction.

On the other hand, \pors are connectionless one-way channels (UDP-like), and the message is not fixed in size due to the lack of re-encryption between each hop.
In particular, \por does not guarantee messages integrity nor order (as each cell potentially travels through a different path).
It is the role of the upper abstraction layer (e.g. our file exchange protocol \spores) to guarantee reliable \& ordered transmission.

%\begin{figure}[ht]
\begin{algorithm}[t]
\caption{The Broadcast Encrypt/Decrypt algorithms}
\label{alg:broadcast_crypto}
%\begin{minipage}[t]{.49\linewidth}
\small
\begin{algorithmic}[1] % [1] means "display all line numbers"
\Function{$BE$}{$\payload, pk_{\layer}$}\label{alg:be}
\State $k \gets $ random symmetric key
\State $\cipher \gets SE(\payload, k)$
\State $\envelope \gets \left\{AE(k, pk)\right\}_{pk \in pk_{\layer}}$
\State \textbf{return} \envelope, \cipher
\EndFunction
\end{algorithmic}
%\end{minipage}\hfill%
%\begin{minipage}[t]{.49\linewidth}

\begin{algorithmic}[1] % [1] means "display all line numbers"
\Function{$BD$}{$\envelope, \cipher, sk$}\label{alg:bd}
\For{$e \in \envelope$}
  \State $k \gets AD(e, sk)$
  \If{$k \neq \bot$}
    \State \textbf{return} $SD(\cipher, k)$
  \EndIf
\EndFor
\State \textbf{return} $\bot$
\EndFunction
\end{algorithmic}
%\end{minipage}%
%
\end{algorithm}

\begin{algorithm}[t]
\caption{The Message Encrypt/Decrypt algorithms}
\label{alg:header_crypto}
%\begin{minipage}[t]{.49\linewidth}
\small

\begin{algorithmic}[1] % [1] means "display all line numbers"
\Function{$ME$}{\payload, \textbf{L}}\label{alg:he}
\State $\mess \leftarrow \payload$
\For{$\layer \in$ \texttt{reverse}(\textbf{L})}
  \State $\mess.\envelope, \mess.\cipher \gets BE(\mess, \layer.pk)$
  \State $\mess.@ \gets \layer.@$
\EndFor
\State \textbf{return} $\mess$
\EndFunction
\end{algorithmic}
%\end{minipage}\hfill%
%
%\begin{minipage}[t]{.49\linewidth}
\begin{algorithmic}[1] % [1] means "display all line numbers"
\Function{$MD$}{\mess, $sk$}\label{alg:hd}

\State \textbf{return} $BD(\mess.\envelope, \mess.\cipher, sk)$

\EndFunction
\end{algorithmic}
%\end{minipage}%

\end{algorithm}

%\end{figure}

\paragraph{Cryptographic primitives}
The encrypted message $\cipher_i$ containing the addresses of the next layer needs to be decipherable by any of the current layer $\layer_i$'s members, and only by them.
This cryptographic scheme is coined Broadcast Encryption (BE)~\cite{fiat_broadcast_1993}.
We derive our encryption process from Hybrid Encryption~\cite{stinson2005cryptography} (as used in PGP), where a message $\mess$ is encrypted into a cipher $\cipher$ using a unique symmetric key $k$ (e.g. using AES).
Each member of the group $\layer_i$ must be given this key, which is the purpose of the envelope $\envelope$.
It contains the concatenation of $k$ encrypted with each member's public key (using e.g. RSA).
Upon reception of a ciphered message $(\envelope, \cipher)$, a peer attempts to decrypt each portion of the envelope with its private key, until it succeeds (and gets $k$ to decrypt $\cipher$) or fails.

We write down our broadcast encryption/decryption algorithms in algorithm~\ref{alg:broadcast_crypto}, and its application to our message cryptography in algorithm~\ref{alg:header_crypto}.

Let $\cipher \gets SE(\payload, k)$ and $\payload \gets SD(\cipher, k)$ be symmetric primitives for en\-crypting/de\-crypting an arbitrary payload \payload with key $k$, such that $SD$ returns $\bot$ on decryption failure. 
% The cipher size $|\cipher|$ is the smallest multiple of the symmetric block size $\symblocksize$ superior or equal to the payload size $|\payload|$. 
%We write that $|\cipher| = \left\lceil \left| \payload \right|\right\rceil_{\symblocksize}$

Let $\cipher \gets AE(\payload, pk)$ and $\payload \gets AD(\cipher, sk)$ be asymmetric ones for encrypting/decrypting a payload \payload with the public key $pk$ (resp. secret key $sk$), such that $AD$ returns $\bot$ on failure.
%The cipher size $|\cipher|$ is always equal to the asymmetric block size \asymblocksize, such that the payload should always be shorter or as long as the block size.

%\noindent
$(\envelope, \cipher) \gets BE(\payload, pk_{\layer})$ --- Given a layer \layer's public keys $pk_\layer$, and an arbitrary payload $\payload$ to encrypt, $BE$ \emph{broadcast encrypts} \payload by outputting an envelope $\mathcal{E}$ containing a symmetric key $k$ encrypted with each $pk \in pk_{\layer}$, and a ciphertext $\mathcal{C}$ containing the payload encrypted with $k$.

%\noindent
$\payload \gets BD(\envelope, \cipher, sk)$ --- Given an envelope $\mathcal{E}$, a ciphertext $\mathcal{C}$ and a secret key $sk$, $BD$ \emph{broadcast decrypts} the payload \payload into the expected plaintext, or $\bot$ if the decryption fails.

In algorithm~\ref{alg:header_crypto}, we write $\layer.pk$ and $\layer.@$ to refer to a layer's nodes' public keys and addresses. 
$\mathbf{L} = \left\{\layer_i\right\}_i$ is an array of layers.
\texttt{reverse}(\textbf{L}) means we iterate on \textbf{L} in reverse order (starting from the last element).

$\mess \gets ME(\payload, \mathbf{L})$ --- Given a payload \payload (that can be a \por message), and an array of layers $\mathbf{L}$, $HE$ recursively encrypts the output message \mess for each $\layer \in \mathbf{L}$ starting from the last.% layer.

$\payload \gets HD(\mess, sk)$ --- $HD$ attempts to decrypt the message \mess using the secret key $sk$. It either returns a payload \payload or $\bot$ on failure.

\begin{algorithm}[t]
\caption{Receiving a \por message  on $d$}
\label{alg:fw}
\small

%\begin{minipage}[t]{.49\linewidth}
\begin{algorithmic}[1] % [1] means "display all line numbers"

\Rec{$\mess$}
  \State $\payload \leftarrow MD(\mess, \secretkey_d)$
  \If{$\payload = \bot$}
    \State $\textbf{return } \bot$ \Comment{Decryption failed}
  \ElsIf{$\payload \neq$ \por message}
    \State $\textbf{process } \payload$ \Comment{I am the recipient}
    %\State $\textbf{return } \top$
  \Else
    \State $\texttt{Forward}(\payload)$
  \EndIf
\EndRec

%\algstore{whatawonderfulname}
\end{algorithmic}
%\end{minipage}\hfill%
%
%\begin{minipage}[t]{.49\linewidth}
\begin{algorithmic}[1] % [1] means "display all line numbers"
%\algrestore{whatawonderfulname}

\Function{$\texttt{Forward}$}{$\mess$}
  \For{$@ \in \texttt{random}(\mess.@)$}
    \If{$\texttt{send}(@, \mess, \timeout)$}\label{alg:fw:send}
      \State $\textbf{return } \top$ \Comment{Success}
    \EndIf
    % \State $\texttt{success} \leftarrow \texttt{send}(@, \langle \header', \payload \rangle, \timeout)$
    % \If{$\texttt{success}$}
    %   \State $\textbf{return } \top$
    % \EndIf
  \EndFor

  \State $\textbf{return } \bot$ \Comment{All layer offline: drop messagge}
\EndFunction

\end{algorithmic}
%\end{minipage}
\end{algorithm}

\paragraph{Forwarding messages}
We finally display the message reception and forwarding procedure in algorithm~\ref{alg:fw}, that runs on any node participating in \spores (we consider a device $d$).
Upon reception of a message $\mess$, $d$ first attempts to decipher it using its secret key $\secretkey_d$.
Three cases are possible: either the decryption fails, which constitutes an error---the message is dropped; either the output is not a \por message, in which case the message is destined to $d$; either $d$ decryption unraveled another message, in which case $d$ forwards it to the next layer using $\texttt{Forward}$ function.
This procedure iterates over each address $@$ in $\mess.@$ \emph{in random order}, and attempts to send the message to $@$.
The $\texttt{send}$ function called at line~\ref{alg:fw:send} takes three parameters: the recipient's address, the message to send, and a timeout duration.
$\timeout$ is a configuration parameter, usually below a second.
If the $\texttt{send}$ call succeeds, the message is duly forwarded.
If the `for' loop returns without any successful attempt, all the next layer is considered offline, and the message is dropped.
Note that it takes $\routesize \times \timeout$ seconds to drop a message when the next layer is offline.

% subsection spor_stateless_probabilistic_onion_routes (end)

\subsection{\spores: File exchanges through \por} % (fold)
\label{sub:spores_file_exchanges_through_spor}

\begin{figure}[t]
	\centering
	\includegraphics[width=0.75\columnwidth]{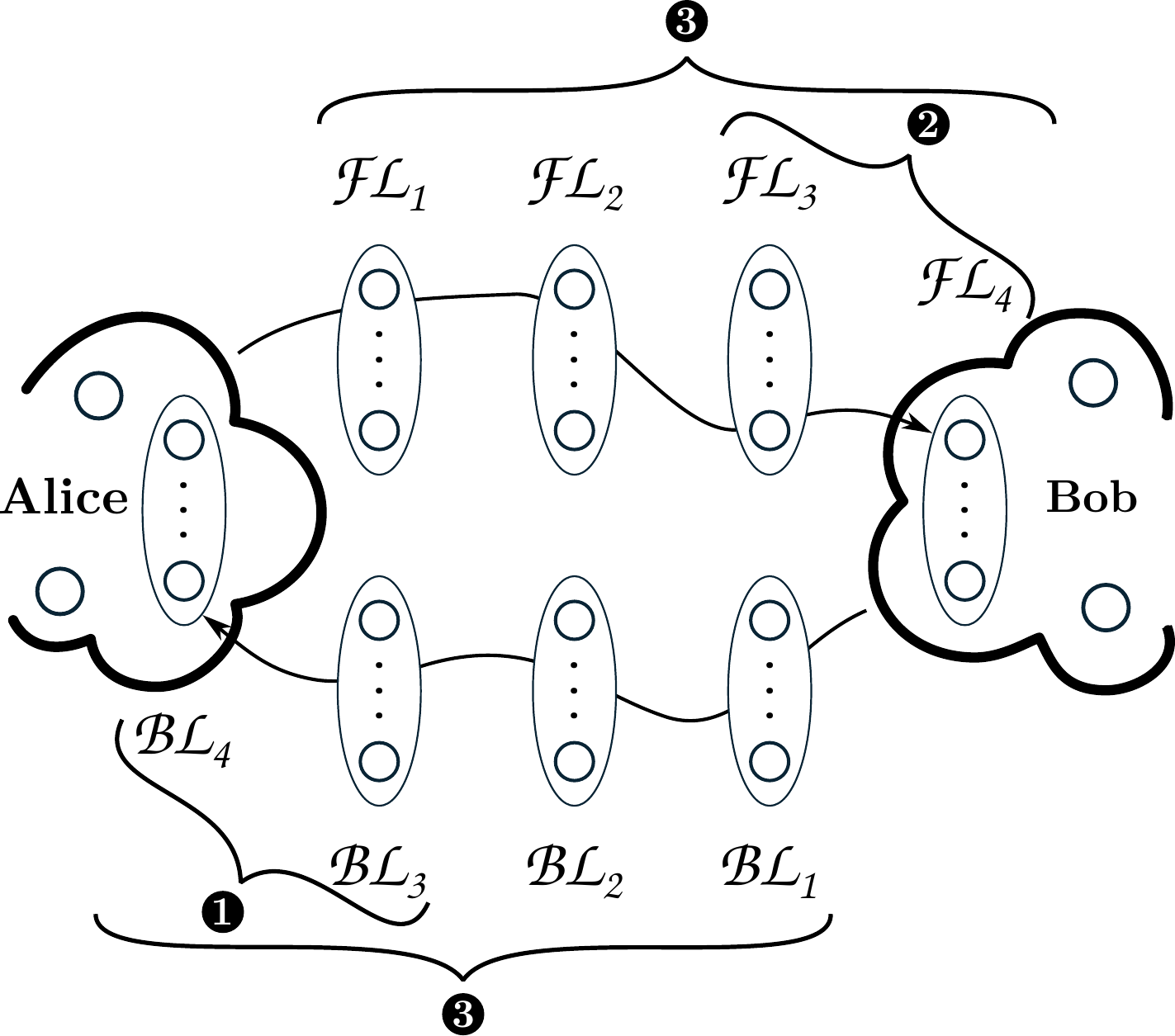}
	\caption{For Alice to send a file to Bob through \spores, they need to agree upon two routes: a forward route to send file chunks, and a backward one to send acknowledgements. Each user contributes layers to each route, so as to maximise the diversity of the involved relays.}
	\label{fig:route_creation}
\end{figure}

We now have all the building blocks to perform anonymous file transfers using e-squads.
In this section, we first present how two users agree upon probabilistic onion routes for their exchange, including the intelligent selection to maximise the routes' availability; finally, we discuss the file exchange protocol built atop \pors.

% \spores participants build \pors to exchange files.
% To ensure sender-receiver anonymity, both parties build a portion of it.

%% I removed the route_creation figure from here

\begin{algorithm*}[h]
\caption{Route initialisation between device $d_A$ uploading file $f$ (on the left) and $d_B$ downloading it (on the right).}
\label{alg:route_creation}

\small
\begin{minipage}[t]{.49\linewidth}
\begin{algorithmic}[1] % [1] means "display all line numbers"

\Function{InitUpload}{$f, \theta$}
  \Statex Step \ding{182}:
  \State $fd \leftarrow \texttt{BuildFileDescriptor}(f)$
  \State $\texttt{sq} \leftarrow \left\{ r.d, r.d \neq d_A \right\}_{r \in S}$ \label{alg:route_creation:d_a:sq}
  \State $\mathcal{BL}_4 \leftarrow \picklayer(\texttt{sq}, \theta) \cup \{ d_A \} $ \label{alg:route_creation:d_a:esquad_layer}
  \State $\mathcal{BL}_3 \leftarrow \picklayer(\vrps, \theta)$
  %\State $\mathcal{BR}_1 \leftarrow HE\left(fd.\texttt{ID}, \left[ \mathcal{BL}_3, \mathcal{BL}_4\right]\right)$ \label{alg:route_creation:innermost}
  \State $\mathcal{BR}_1 \leftarrow \left[ \mathcal{BL}_3, \mathcal{BL}_4\right]$ \label{alg:route_creation:d_a:innermost}
  \State send $\langle fd, \mathcal{BR}_1 \rangle$ to $d_B \hfill \longrightarrow\;\longrightarrow\;\longrightarrow\;\longrightarrow$ \label{alg:route_creation:d_a:send}
\EndFunction

\Statex
\Statex
\Statex
\Statex
\vspace{2ex}

\setcounter{ALG@line}{11}
\Rec{$\langle \mathcal{FR}_1 \rangle$} $\hfill \longleftarrow\;\longleftarrow\;\longleftarrow\;\longleftarrow$
  \Statex Step \ding{184}:
  % Finish forward header
  \State $\mathcal{FL}_1 \leftarrow \picklayer(\vrps, \theta)$ \label{alg:route_creation:layer_1} 
  \State $\mathcal{FL}_2 \leftarrow \picklayer(\vrps, \theta)$
  %\State $\mathcal{FR} \leftarrow HE\left(\mathcal{FR}_1, \left[ \mathcal{FL}_1, \mathcal{FL}_2\right]\right)$ \label{alg:route_creation:finish_route}
  \State $\mathcal{FR} \leftarrow \left[ \mathcal{FL}_1, \mathcal{FL}_2\right] \oplus \mathcal{FR}_1$ \label{alg:route_creation:finish_route}
  % Share the new file exchange
  \State $r \leftarrow (\texttt{Now}(), d_A, \text{UL}, fd.\texttt{ID})$ \label{alg:route_creation:round}
  \State $S \leftarrow S \cup \{r\}$ \Comment{Shared to e-squad} \label{alg:route_creation:seq}
  % Start sending
  \State Start sending $f$ \label{alg:route_creation:start}
\EndRec

\end{algorithmic}
%\end{algorithm}
\end{minipage}\hfill%
\begin{minipage}[t]{.49\linewidth}
%\begin{algorithm}[H]
\begin{algorithmic}[1] % [1] means "display all line numbers"
%\caption{Route initialisation by device $d_B$ downloading file $f$}

\Statex
\Statex
\Statex
\Statex
\Statex
\Statex
\Statex
%\vspace{1.5ex}
\setcounter{ALG@line}{6}
\Rec{$\langle fd, \mathcal{BR}_1 \rangle$} \label{alg:route_creation:rec}
  \Statex Step \ding{183}:
  % Build forward header
  \State $\texttt{sq} \leftarrow \left\{ r.d, r.d \neq d_B \right\}_{r \in S}$ \label{alg:route_creation:sq}
  \State $\mathcal{FL}_4 \leftarrow \picklayer(\texttt{sq}, \theta) \cup \{ d_B \}$  \label{alg:route_creation:esquad_layer}
  \State $\mathcal{FL}_3 \leftarrow \picklayer(\vrps, \theta)$
  %\State $\mathcal{FR}_1 \leftarrow HE\left(fd.\texttt{ID}, \left[ \mathcal{FL}_3, \mathcal{FL}_4\right]\right)$ \label{alg:route_creation:innermost}
  \State $\mathcal{FR}_1 \leftarrow \left[ \mathcal{FL}_3, \mathcal{FL}_4 \right]$ \label{alg:route_creation:innermost}
  \State send $\langle \mathcal{FR}_1 \rangle$ to $d_A$ \label{alg:route_creation:send}
  % Finish backward header
  \Statex Step \ding{184}:
  \State $\mathcal{BL}_1 \leftarrow \picklayer(\vrps, \theta)$ \label{alg:route_creation:layer_1}
  \State $\mathcal{BL}_2 \leftarrow \picklayer(\vrps, \theta)$ 
  %\State $\mathcal{BR} \leftarrow HE\left(\mathcal{BR}_1, \left[ \mathcal{BL}_1,  \mathcal{BL}_2\right]\right)$ \label{alg:route_creation:finish_route}
  \State $\mathcal{BR} \leftarrow \left[ \mathcal{BL}_1,  \mathcal{BL}_2\right] \oplus \mathcal{BR}_1$ \label{alg:route_creation:finish_route}
  % Share the new file exchange
  \State $r \leftarrow (\texttt{Now}(), d_B, \text{DL}, fd.\texttt{ID})$ \label{alg:route_creation:round}
  \State $S \leftarrow S \cup \{r\}$ \Comment{Shared to e-squad} \label{alg:route_creation:seq}
  \State Start receiving $f$ \label{alg:route_creation:start}
\EndRec

\end{algorithmic}
\end{minipage}
\end{algorithm*}

%\vspace{2ex}

\paragraph{Routes creation} Fig.~\ref{fig:route_creation} depicts the creation process of a route between our beloved Alice (uploader) and Bob (the receiver).
As already mentioned, this process takes place out-of-band (on another communication channel such as Near Field Communication (NFC), LAN, Bluetooth, carrier pigeon, or else).
%As already mentioned, we leverage on proxemics relationships, such that the handshaking process initialising a file exchange is performed out-of-band (on another communication channel such as Near Field Communication (NFC), LAN, Bluetooth, carrier pigeon, or else).
The initialisation serves two purposes: to provide Bob with the exchanged file metadata (we come back to it in the next paragraphs), and to decide upon the \pors that will be used throughout the transfer.

Since \pors are one-way only, Alice and Bob need to agree upon two routes: a \emph{forward} one, form Alice to Bob, that will carry file chunks, and a \emph{backward} route, from Bob to Alice, that will transport Bob's acknowledgements of the chunks. 
Furthermore, to maximize the peer diversity in the route (for security reasons), both parties compute a portion of each route.
In the rest of the paper, we settle with a number of layers of $\routesize=4$, which is required to have decent security properties, while adding more layers would not make routes significantly more secure (as agreed upon by most onion routing approaches).
% Furthermore, as \spores heralds sender-receiver anonymity, both parties build a portion of the route, so as to maintain their end of the route hidden inside the encrypted header.

We now detail the operations depicted in Fig.~\ref{fig:route_creation}, and detailed in algorithm~\ref{alg:route_creation}:
\begin{itemize}
\item At \ding{182}, Alice's sending device $d_A$ crafts the inner part of the header for the route to herself, $\backwardroute_1$, at lines~\ref{alg:route_creation:d_a:sq}-\ref{alg:route_creation:d_a:innermost}.
	The final layer $\backwardlayer_4$ is only constituted of Alice's devices:
	$d_A$ picks candidates from her e-squad sequence $S$ (line~\ref{alg:route_creation:d_a:sq}), 
	and lastly adds its own descriptor to $\backwardlayer_4$ (l.~\ref{alg:route_creation:d_a:esquad_layer}).
	The third layer $\backwardlayer_3$ is selected using $d_A$'s RPS view \vrps, which contains a pool of global descriptors.
	On line~\ref{alg:route_creation:d_a:send}, $d_A$ sends $\backwardroute_1$ to Bob, along with the file metadata $fd$.

\item At \ding{183}, Bob's receiving device $d_B$ builds its half of the forward route ($\forwardroute_1$) just like $d_A$ did at \ding{182}, see lines~\ref{alg:route_creation:sq}-\ref{alg:route_creation:innermost}.
	Again, $\forwardlayer_4$ is only made of Bob's e-squad, while $\forwardlayer_3$ samples devices from the global overlay.
	$d_B$ sends $\forwardroute_1$ back to $d_A$ on line~\ref{alg:route_creation:send}.

	% Bob also finalises the backward route: 
	% $\backwardroute = HE(\backwardroute_1, [\backwardlayer_1])$ 
	% and sends $\forwardroute_1$ back to Alice (but keeps \backwardroute to himself).

\item Finally, at \ding{184}, both devices bootstrapping before starting the file exchange.
	They first finish the route they will use to reach the other end (lines~\ref{alg:route_creation:layer_1}-\ref{alg:route_creation:finish_route}), 
	then inform their e-squad that they started sharing a file by adding an interaction $r$ to their sequence $S$ (lines~\ref{alg:route_creation:round}-\ref{alg:route_creation:seq}),
 	and finally start exchanging $f$ (line~\ref{alg:route_creation:start}).
% Alice simply finishes the forward route:
% 	$\forwardroute = HE(\forwardroute_1, [\forwardlayer_1])$ 
% 	She can now start sending her file to Bob, that will be able to answer with acknowledgements on the backward route.
\end{itemize}

\paragraph{Relays selection} 
We now detail the $\picklayer(\mathcal{V}, \theta)$ function, that takes care of intelligently selecting a layer's devices. 
It takes two parameters: an input set of candidate nodes $\mathcal{V}$, and the \emph{unavailability threshold} $\theta\in \left] 0, 1\right]$, a configuration parameter that represents the desired maximum probability that all of the layer's nodes fall offline at the same time (i.e. the probability that the layer be unavailable).

\picklayer iteratively picks a \emph{random} node from $\mathcal{V}$ without replacement, adds it to the output layer \layer, and computes the \emph{probability that all of the layer's nodes fall offline at once}, $\Poff_\layer$:
\begin{equation}
\Poff_\layer = \prod\limits_{d \in \layer}{ 1 - P_i(d) }
\label{eq:poff}
\end{equation}
$P_i(d)$ being the probability that device $d$ remains online (cf. Eq.~\ref{eq:p_d_markov}).
The function returns either when the offline probability $\Poff_\layer$ falls below the threshold $\theta$, or when the input view $\mathcal{V}$ is emptied.
As a baseline that will be used in the evaluation, \picklayer  randomly chooses a predetermined number of nodes from the input view, without caring for the layer's probability of becoming unavailable.

Non-e-squad layers are built with \vrps as input: it comprises a random pool of global \spores participants that were online not long ago at least (cf. Sec.~\ref{sub:the_global_overlay}).
The RPS view size \viewsize should be chosen big enough for \picklayer to reach the configured $\theta$, but small enough that the view's stale descriptors get evicted in a reasonable amount of time.
%Usually, \picklayer is able to reach $\theta$ before \vrps is emptied, as the view size \viewsize is chosen .
E-squad layers, on the other hand (that is, $\forwardlayer_4$ and $\backwardlayer_4$), only have the less numerous e-squad members as input, such that \picklayer might not be able to reach the threshold before emptying the candidate list.

The smaller the threshold $\theta$, the more nodes per layer, the better the route's availability, but also the bigger the header.
There is a trade-off between the readiness of routes and the message transit overhead.

Finally note that randomly picking descriptor from one's view avoids biasing the relay selection in favour of supposedly highly connected nodes.
Indeed, the devices' availability estimate is published by themselves, and should not be trusted. 
Our approach gives no interest for attackers to lie on this value, while it encourages everyone to provide good estimates, for the sake of the routes' reliability.

\paragraph{Exchanging a file} 
As already told, \por provides anonymous UDP-like channels: order and integrity of the messages are not guaranteed by the protocol.
These features must be supplied by \spores on top of \por.%---in our case: \spores.

A file $f$ exchanged through \spores is chunked into fixed-size pieces, that are transmitted in order by the sender, along with their position (or ID).
To ensure chunks integrity, we borrow from BitTorrent~\cite{bep3}:
the file descriptor $fd$ that is computed with $\texttt{BuildFileDescriptor}(f)$ and provided to the receiver on bootstrap notably contains a \texttt{SHA1} hash per chunk.
The receiver verifies that the expected and computed hashes match every time they receive a chunk. 
The function creates the following descriptor:
$$ fd = \left( \texttt{ID}, \texttt{size}, \texttt{chunkSize}, \texttt{\#Chunks}, \texttt{chunksHash}, \texttt{hash}\right) $$
Each file is given a unique, random $\texttt{ID}$, picked by the uploader.
The file descriptor also provides the file $\texttt{size}$, number of chunks and chunk size. 
The $\texttt{chunksHash}$ is the concatenation of each chunk's SHA1 hash, used by the receiver to verify the integrity of each chunk.
Finally, $\texttt{hash}$ is the SHA1 hash of $\texttt{chunksHash}$, to verify its own integrity.
Using SHA1 hashes, we ensure the file integrity. The order is guaranteed by the following sliding-window protocol.

To accelerate the file exchange, \spores implements the Selective Repeat Automatic Repeat-reQuest (ARQ)~\cite{lockitt_selective_1975,weldon_improved_1982,peterson_computer_2003} algorithm, a sliding-window protocol that lets the sender send several chunks at once, and allows the receiver to accept them out of order.
The sender provides the chunk ID of each piece sent on the forward route, while the receiver sends back an acknowledgement (ACK) with the same ID for each received piece, using the backward route.
%The receiver sends an acknowledgement (ACK) back to the sender with the ID for every received chunk, even if it was already acknowledged.
When the sender does not receive an ACK after sending a chunk, it retries sending after a timeout of several seconds.
The file exchange completes once each file chunk has been ACKed.

Finally, as can be seen in lines~\ref{alg:route_creation:d_a:esquad_layer} and \ref{alg:route_creation:esquad_layer} of algorithm~\ref{alg:route_creation}, any e-squad member can receive chunks/ACKs in spite of the proper message recipient.
When they do, they can unravel the payload, and forward it to its proper recipient, by finding the recipient's address in their e-squad sequence $S$.
%By comparing the innermost part of the receiver header (cf. alg.~\ref{alg:route_creation} line~\ref{alg:route_creation:innermost} and alg.~\ref{alg:route_creation} line~\ref{alg:route_creation:innermost}) to the file IDs in their sequence $S$, each e-squad member knows to which device a message is destined, and can forward it to their recipient.
If the receiver is currently offline, they forward the message to any online e-squad member, until the recipient comes back online and is able to finally receive the message.
In essence, the whole e-squad acts as a cache for received messages while the actual recipient is offline.

% Indeed, we already saw that a subset of one's e-squad constitute the final layer of the forward and backward routes (see $\forwardlayer_3$ and $\backwardlayer_3$ in Fig.~\ref{fig:route_creation}).
% When a device receives a message that is destined for a fellow e-squad member, it intelligently attempts to forward it to its proper recipient by looking into their user activity sequence: may it be disconnected, it waits for the recipient to come back online before handing over the message.

% subsection spores_file_exchanges_through_spor (end)

\medskip
With these building blocks, we have proposed an entirely decentralised anonymous file exchange service for e-squads. 
It is specifically tailored for networks with high churn, and, thanks to its gossip components, it can scale to a theoretically unbounded number of users.
We now analyse the security properties of \spores, before evaluating its prototype.

% section spores_approach (end)

\section{Security analysis} % (fold)
\label{sec:attack_model}

We claim that using \spores for exchanging files is more anonymous than using traditional onion routing (e.g. OnionShare on Tor~\cite{dingledine_tor_2004}).
To this aim, we statistically compare \spores' and Tor's resilience to de-anonymisation attacks.

\subsection{Assumptions and threat model} % (fold)
\label{sub:assumptions_and_threat_model}

We take interest in an attacker owning a portion of the network relays (at the very least, their e-squad), and that can tamper with the protocol's specification (they can notably break the random selection while forwarding messages in alg.~\ref{alg:fw}).
Their goal is to link two file exchange participants.

We do not consider the infamous Global Passive Adversary (GPA) attack model, where an attacker listens on all communication pipes.
Using people's devices drastically increases the number of Autonomous Systems (AS) involved in the protocol (mobile carriers, household connections...), rendering the GPA unlikely.
%We deem this attack impractical due to the number of Autonomous Systems (AS) involved \commentAL{mobile carriers, household connections...}.
In any case, GPA circumvention almost always involves generation of cover traffic~\cite{Freedman:2002,van_den_hooff_vuvuzela_2015,piotrowska_loopix_2017,podolanko_lilac_2017}, which we cannot afford on constrained user end-devices.

Because we leverage secure peer sampling~\cite{jesi_secure_2010}, we assume that the global overlay cannot be tampered with, and does return a uniform sample of online peers in the system.
Although headers are of variable size, we further assume that relays cannot guess their position on a route, as they do not know the number of relays per layer.

We already stated that the number of online relays in Tor ($\sim$6000) was small relative to the number of connected users ($\sim$2 million).
\spores seeks to involve each user device as a relay.
Hence, we make the assumption that the number \nspores of relays in \spores is a multiple of those of Tor: $\nspores = C \times \ntor$, with $C \ge 1$.
We further assume that there are \nadv colluding devices trying to de-anonymise Alice and Bob as they exchange a file.
We write $\pspores = \nadv/\nspores$ the proportion in attackers in \spores, and $\ptor = \nadv / \ntor$ the one in Tor.
Finally, we consider $\routesize=3$ hops per route (excluding the final layer composed only of the recipient's e-squad), and we assume a constant amount of \layersize relays per layer.

\subsection{Likelihood of the traffic correlation attack} % (fold)
\label{sub:likelihood_of_end_to_end_traffic_correlation}

It is well established that onion routing and Tor in particular are not resilient to end-to-end traffic correlation attacks~\cite{dingledine_tor_2004,serjantov_passive_2003,johnson_users_2013,rochet_waterfilling_2017}. 
An attacker listening to each end of an onion route (by owning both end relays or observing traffic) can easily link sender and receiver, and thus de-anonymise the connection.
In \spores, due to the several relays per hop, all messages do not follow the same path.
Considering also the increased number of relays in \spores, we claim that traffic correlation attacks are more difficult than in Tor.

We do not model the operation of the traffic correlation attack.
Instead, we study the probability that an adversary successfully positions themself on a route's first and last hops, and receives transmitted messages on both ends.
We call this overall probability $P\left[ \text{analyse mess.} \right]$.

\paragraph{In Tor,} the probability of having an adversary observe the same message on the first and last hops---knowing that they own these relays---is 1, since messages all go through the same relays once the route is built.
In other words, the probability of seizing messages in Tor, $P^\textsc{Tor}\left[ \text{analyse mess.} \right]$, is simply the probability that the adversary successfully positions themself on the first and last hop.
We assume that the probability $P^\textsc{Tor}\left[ \text{pick adv.} \right]$ of selecting an adversary is the same for each layer (an overestimation of Tor's actual security), and that relays are selected with replacement (which eases the computation, while only having a negligible impact on the outcome).
Under these terms: 
$$
P^\textsc{Tor}\left[ \text{analyse mess.} \right] = P^\textsc{Tor}\left[ \text{pick adv.} \right]^2 = \left(\frac{\nadv}{\ntor}\right)^2
$$
\paragraph{In \spores,} it gets more complex.
On the first and last layers, the situation is the same: the adversary has to own a number $k$ of relays in the layer of size \layersize; then, the previous layer has to forward the message to the adversarial nodes.
We consider both cases to have the same independent probability: 
\begin{multline*}
P^\textsc{Sp}\left[ \text{adv. reads mess.} \in \layer  \right] = \sum\limits_{k=1}^{\layersize} P^\textsc{Sp}\left[ k \text{ adv.} \in \layer \right] \times \\
P^\textsc{Sp}\left[ \text{adv. reads mess.} \in \layer \mid k \text{ adv.} \in \layer \right]
\end{multline*}
Applying the standard equation for sampling with replacement, and considering a uniform probability of picking an adversary when they are $k$ among \layersize, we obtain:
\begin{align*}
&P^\textsc{Sp}\left[ \text{adv. reads mess.} \in \layer  \right] = \sum\limits_{k=1}^{\layersize} \binom{\layersize}{k} \; \pspores^k \left( 1 - \pspores \right)^{\layersize - k} \times \frac{k}{\layersize} \\
%&\mathop{=\joinrel=}_{j=k-1}^{M=\layersize-1} 
&=\pspores \sum\limits_{j=0}^{M} \binom{M}{j} \; \pspores^j \left( 1 - \pspores \right)^{M  - j} 
= \pspores \left( \pspores + \left( 1 - \pspores\right)\right)^M = \pspores \\
%\vspace{10pt}&\\
&\implies P^\textsc{Sp}\left[ \text{analyse mess.} \right] = P^\textsc{Sp}\left[ \text{adv. reads mess.} \in \layer  \right]^2=\pspores^2
\end{align*}
We see that, in \spores, the lesser security of selecting more nodes per layer (thus augmenting the probability to pick an adversary per hop) is strictly compensated by the probability to send a message to the adversaries.
The probability that an adversary reads a message on a layer is equivalent to the probability of picking an adversary.
\paragraph{Comparing the approaches}
We see that \spores' probability of of traffic analysis is better than Tor's as long as $C$ is superior to one (that is, as long as there are more relays in \spores than in Tor):
\begin{align*}
P^\textsc{Sp}\left[ \text{analyse mess.} \right] &< P^\textsc{Tor}\left[ \text{analyse mess.} \right]  \\
\iff \left( \frac{\nadv}{C \times \ntor} \right)^2 &< \left( \frac{\nadv}{\ntor} \right)^2 \iff C > 1
\end{align*}
Since \spores is specifically tailored to enable low-end client devices to participate in the network, which would increase the number of relays in the network, deploying probabilistic onion routes on a legacy network like Tor would indeed improve security.
% subsection likelihood_of_end_to_end_traffic_correlation (end)

\subsection{Likelihood of having adversaries on each hop} % (fold)
\label{sub:likelihood_of_having_adversaries_on_each_hop}

Tor does not take much interest in the probability that adversaries own all relays on a circuit (trivially de-anonymising the route), as it is negligible with regards to to the probability that they perform traffic correlation attacks.
Still, because \spores selects several nodes per layer, and because adversaries could break the random selection of relays while forwarding messages (and intently pick their accomplices in the next layer until destination), this attack vector needs to be studied in our case.
We note this attack's probability $P\left[ \forall i, \text{adv} \in \layer_i \right]$.

\paragraph{In Tor's case,} still considering that each layer's probability of picking an adversary is independent and equal, the probability that adversaries own the whole route is simply: 
$P^\textsc{Tor}\left[ \forall i, \text{adv} \in \layer_i \right] = \left(\ptor\right)^{\routesize}$.

\paragraph{\spores' case}\hspace{-0.5em}is again more complex.
The adversary must first receive a message on the first layer $\layer_1$, i.e. with a probability of $P^\textsc{Sp}\left[ \text{adv. reads mess.} \in \layer  \right]=\pspores$.
Then, they must own at least one relay on each of the next layers $\layer_2$ and $\layer_3$, which for each layer has the probability:
\begin{align*} 
&P^\textsc{Sp}\left[ \text{adv} \in \layer \right] = 1 - P^\textsc{Sp}\left[ \text{adv} \notin \layer \right] = 1 - \left(1 - \pspores \right)^{\layersize} \\
&\mathop{=\joinrel=}_{q=1-\pspores} \left( 1 - q \right) \times \sum\limits_{k=0}^{\layersize-1} q^k = \pspores \times \sum\limits_{k=0}^{\layersize-1} \left(1 - \pspores \right)^k
%=\; &\pspores \times \sum\limits_k^{\layersize} (1 - \pspores)^k
\end{align*}
The probability that the adversary owns relays on each layers and successfully forwards it from source to destination is then:
\begin{align*}
P^\textsc{Sp}\left[ \forall i, \text{adv} \in \layer_i \right] &= P^\textsc{Sp}\left[ \text{adv. reads mess.} \in \layer_1  \right] \times P^\textsc{Sp}\left[ \text{adv} \in \layer \right]^2 \\
&= \pspores^3 \times \left( \sum\limits_{k=0}^{\layersize-1} \left(1 - \pspores \right)^k \right)^2
\end{align*}

\paragraph{Comparing the approaches}
If we overrate $\left(1 - \pspores \right)^k \approx 1$, we get the following inequality:

\begin{align*}
&P^\textsc{Sp}\left[ \forall i, \text{adv} \in \layer_i \right] < P^\textsc{Tor}\left[ \forall i, \text{adv} \in \layer_i \right]  \\
&\approx \left( \frac{\nadv}{C \times \ntor} \right)^3 \times \layersize^2 < \left( \frac{\nadv}{\ntor} \right)^3 \implies C > {\layersize}^{2/3}
% \left( \frac{\nadv}{C \times \ntor} \right)^2 &< \left( \frac{\nadv}{\ntor} \right)^2 \iff C > 1
\end{align*}

Considering that increasing the layer size \layersize past a certain threshold yields no performance gain (see section~\ref{ssub:tuning_probabilistic_onion_routes}), and is costly in terms of message size and transmission time, we recommend values strictly inferior to $\layersize=20$.
With this upper bound, $C=7.4$.
We do expect a deployment of \spores, with the same userbase as Tor, to reach a much bigger number of relays than $7.8 \times \ntor = 46800$.

\medskip
In this section, we have seen that \spores' churn-resilient onion routing approach---the multi-path \pors---was not detrimental to its security under the two attack scenarii that we covered. 
In fact, assuming a bigger amount of relays than in Tor (i.e. assuming that Tor implements \pors), probabilistic onion routing even yields a security improvement.

\section{Evaluation} % (fold)
\label{sec:evaluation}

In this section, we evaluate \spores in terms of privacy and performance, depending on the user behaviour, and comparing to existing proposals.
We first describe our evaluation protocol, before presenting our results in section~\ref{sub:conducted_experiments}.

\subsection{Testbed} % (fold)
\label{sub:testbed}

Let us first present how we simulated user behaviours, before going through our experimental setup.

\subsubsection{User behavioural models} % (fold)
\label{ssub:users_behavioural_models}

To the best of our knowledge, there exists no dataset that we could use to represent the behaviour of an e-squad owner.
For this reason, we propose several models for simulating users, with the objective of generating devices' connection and disconnection patterns encompassing the complexity of human behaviour.

We employ a discrete-time Hidden Markov Model (HMM) of order one~\cite{rabiner_introduction_1986} to represent a user going to different places (hidden process), and their device usage patterns depending on their location (observable processes).
We thus assume that users switch location with a fixed period of $T$ seconds, 
and that their next location only depends on the previous one (the Markovian hypothesis).
Each device is modelled by a independent process: their availability only depends on the user's location (and not on other devices).
The concept of availability encapsulates both the power state and transient connectivity of devices.

\paragraph{An example} To illustrate our model, we display a fictitious user model comprising $\nlocations=3$ different locations and $\ndevices=4$ devices.
Below are represented the $\nlocations \times \nlocations$ matrix $A$, that drives the user's movements, and the $\nlocations \times \ndevices$ matrix $B$, that is the concatenation of each device's probability at each state (i.e. nothing sums to one).
Note that nothing prevents the user from using several devices at a time ($B$'s rows do not sum to one):

\scriptsize
$$ A = 
\kbordermatrix{
      		& Home 	& Outside	& Work	\cr
    Home 	& 0.6 	& 0.4 		& 0 	\cr
    Outside & 0.2 	& 0.6 		& 0.2 	\cr
    Work 	& 0 	& 0.4  		& 0.6  	\\[0.3em]
}$$

$$ B = 
\kbordermatrix{
      		& Phone & Laptop& Home~computer & Workstation \cr
    Home 	& 0.8 	& 0.6 	& 0.7			& 0 		\cr
    Outside	& 0.6 	& 0.2 	& 0 			& 0			\cr
    Work	& 0.7 	& 0.2	& 0 			& 0.7 		\\[0.3em]
}$$
\normalsize

% We see that this user cannot go to work without going outside; that they use their phone and laptop with a variable probability depending on the location;
% that the home computer is only accessed at home, and the workstation at work.
% Nothing prevents the user from using several devices at the same time.

Once a model is built, we perform a random walk of $L$ rounds to generate a sequence, or timeline, of interactions $X \in \left\{ 0, 1\right\}^{\ndevices \times L}$.
Like in section~\ref{ssub:modelling_the_user_s_behaviour}, $X_i(d) = 1$ means that device $d$ was online at round $i$, and equals 0 otherwise.
%Figure~\ref{fig:hmm_timeline} shows such a timeline sampled from the previous example.
Devices, through the e-squad overlay, learn the timeline $X$ to predict their future availability, but not the hidden location sequence of their user.

% \begin{figure}[h]
% 	\centering
% 	\includegraphics[width=\columnwidth]{figures/hmm_timeline.pdf}
% 	\caption{A possible device connection timeline output by a random walk on the above matrices $A$ and $B$.}
% 	\label{fig:hmm_timeline}
% \end{figure}

\paragraph{Diverse user models}
With this HMM ground, we can build models with variable mean availability and predictability.
Consider, for instance, a model with only one location, with every device's probability equal to 0.5. Intuitively, this model is the most unpredictable we could build, as all devices switch state with uniform probability.
On the contrary, a model with $L$ locations that are visited in order by the user, and devices probabilities that are either 1 or 0, is very predictable: the device connection timeline is deterministic, and loops every $L$ usage rounds.

To build matrices $A$ and $B$ that display such diversity, we sample their content using the beta distribution, a versatile probability distribution function defined on $[0, 1]$, first studied by Pearson in 1895~\cite{pearson_x._1895}.
The matrix $A$ is then normalized as needs be.
The beta distribution has two shape parameters $(\alpha, \beta) \in R^{+*}$; we are interested in the function's smoothed binomial shape when both parameters are below one.
Small values of $\alpha=\beta$ lead to samples closer to 0 or 1, while $\alpha=\beta=1$ is the uniform distribution.
We skew the distribution, for a fixed $\beta=0.6$, by varying the expected value $\mu$ by picking $\alpha=\frac{\beta}{\mu^{-1} - 1}$. 

% Figure~\ref{fig:beta_dist} shows, on top, how small values of $\alpha=\beta$ lead to samples closer to 0 or 1, while $\alpha=\beta=1$ is the uniform distribution.
% On the bottom, we skew the distribution, for a fixed $\beta=0.6$, by varying the expected value $\mu$ by picking $\alpha=\frac{\beta}{\mu^{-1} - 1}$.

We measure the predictability of a given model by generating a timeline $X$ of length $L$.
To evaluate the predictability, we first compute the probability $P_i(d)$ that each device $d$ stays online, for each round $X_i$ such that $L_\text{init}<i<L$.
(Since computing $P_i(d)$ requires initial information, we only compute it for steps past a number of rounds $L_\text{init}$.)
Then, we compare $P_i(d)$ with the actual outcome $X_{i+1}(d)$ using a logarithmic scoring rule, that is: $\text{sc}_i(d) = X_{i+1}(d) log(P_i(d)) + (1-X_{i+1}(d)) log(1-P_i(d))$.
Finally, the total predictability of a model is the average of all computed log scores.

% We measure mean availability, churn rate and predictability for a given model by generating a timeline $X$ of length $L$.
% The mean availability is straightforward to measure: a simple average of the number of times devices were online over the whole timeline.
% We employ Godfrey et al.'s churn metric~\cite{godfrey_minimizing_2006}: it is the average, over each usage round, of the fraction of devices that changed state at this round (it can be superior to one).
% To evaluate the predictability, we first compute the probability $P_i(d)$ that each device $d$ stays online, for each round $X_i$ such that $L_\text{init}<i<L$.
% (Since computing $P_i(d)$ requires initial information, we only compute it for steps past a number of rounds $L_\text{init}$.)
% Then, we compare $P_i(d)$ with the actual outcome $X_{i+1}(d)$ using a logarithmic scoring rule, that is: $\text{sc}_i(d) = X_{i+1}(d) log(P_i(d)) + (1-X_{i+1}(d)) log(1-P_i(d))$.
% Finally, the total predictability is the average of all computed log scores.

We proposed 4 user behavioural models representing diverse predictabilities.
For each of them, unless otherwise noted, we set $\nlocations=4$ locations and $\ndevices=6$ devices. % and a mean availability of $\mu=0.5$.
$L_\text{init}$ is always set to 50, while the total sequence length $L$ depends on the experiment duration:

\begin{enumerate}
	\item Uniform (\emph{Uni.}): $\nlocations=1$; each device's probability is equal to $\mu$. This model shows no periodicity, and is thus the least predictable.
	%\item Beta Uniform (\emph{Beta Uni.}): $A$ and $B$ values are sampled from a beta distribution with $\beta=1$ and the configured $\mu$. The output HMM is very unpredictable.
	\item Unpredictable (\emph{Unpred.}): as above, the HMM matrices are sampled from a beta distribution, this time with $\beta=0.8$, generating transitions and device probabilities closer to 0 or 1.
	\item Predictable (\emph{Pred.}): here, $\beta=0.2$, which brings probabilities even closer to 0 or 1.
	%\item Beta Deterministic (\emph{Beta Det.}): with $\beta=0.0001$, output probabilities are \emph{always} equal to 0 or 1 with a negligible epsilon.
	\item Deterministic (\emph{Det.}): the user cycles deterministically through the set of $\nlocations$ locations. Devices probability being always 0 or 1, the timeline $X$ is entirely deterministic.
\end{enumerate}

% \begin{figure}[t]
% 	\centering
% 	\includegraphics[width=.7\columnwidth]{figures/beta-dist_landscape.pdf}
% 	\caption{Probability density function of the beta distribution for several parameter values.
% 	On top, we see that when $\alpha=\beta$ are small, the function produces values closer to 0 or 1 (binomial). $\alpha=\beta=1$ is the uniform function.
% 	On the bottom, we fix $\beta=0.6$, and vary $\alpha$ such as to change the function's expected value $\mu$.}
% 	\label{fig:beta_dist}
% \end{figure}

We evaluate the performance of \spores with regard to these different models in section~\ref{ssub:influence_of_the_user_s_behaviour}.

% subsubsection users_behavioural_models (end)

\subsubsection{Methodology} % (fold)
\label{ssub:methodology}

To evaluate \spores, we built a prototype in 6100 lines of Go, including all core functions except the cryptography.
The users' behaviours, driving the devices churn, were simulated with 1600 lines of Python.
Each device runs as a Docker container, participating in a single virtual network.
Due to the scale of the experiment, and to generate somewhat realistic network traffic, each user's devices are scattered over a multi-host Docker Swarm.
The experiments were deployed on 6 AWS `r5.large' VMs, plus another one to orchestrate the experiments.

The experimental process is the following: 
we initialise the experiment by letting each user's device spawn on a random VM.
At this time, booted devices start exchanging descriptors in the global overlay.
Once every device is started, we start scheduling each device according to their user's behavioural model, updating their availability state every $T$ seconds.
We then pick two random online devices belonging to different users, and perform the initial route creation through REST calls to each of them.
We repeat the operation until we consider enough files were exchanged, leaving a reasonable time between exchanges to avoid saturating the network.
We tear down the network and retrieve results after leaving some time for the devices to proceed with their file exchanges.
Note that the orchestrating VM is not able to assess whether files finished exchanging, such that some files fail downloading for lack of time.

\paragraph{Parameters}
There are $\nusers=25$ users in the system, each running 6 devices, resulting in a network of 150 relays, randomly scattered over the hosts.
Users switch between $\nlocations=4$ states.

We fixed the file size to 50MiB, and the chunk size to 512KiB, resulting in 100 chunks per file transfer.
To bootstrap the e-squad overlay, we provide them with an initial user activity sequence of $L_\text{init}=50$ device usages. 
This way, even the initial availability predictions are backed by a reasonably accurate model.

Each user interaction lasts $T=6$ seconds; we leave $5*T=30s$ between each file exchange; exchange 50 files per experiment, and tear down the experiment $20*T=2m$ after the last transfer started.

Unless otherwise noted, the unavailability threshold $\theta$ for creating routes equals 0.001.
When the user model is not specified, the unpredictable is under study.
When the mean availability $\mu$ is not specified, it equals 50\%.

% subsubsection methodology (end)

% subsection testbed (end)

% \begin{figure}[t]
% 	\centering
% 	\includegraphics[width=0.8\columnwidth]{figures/success_rate_f_mean_avail.pdf}
% 	\caption{File exchange success rate for each model, as a function of the mean availability $\mu$. The grey dotted line represents $x=y$.}
% 	\label{fig:success_rate_f_avail}
% \end{figure}

\begin{figure}[t]
	\centering
	\includegraphics[width=0.8\columnwidth]{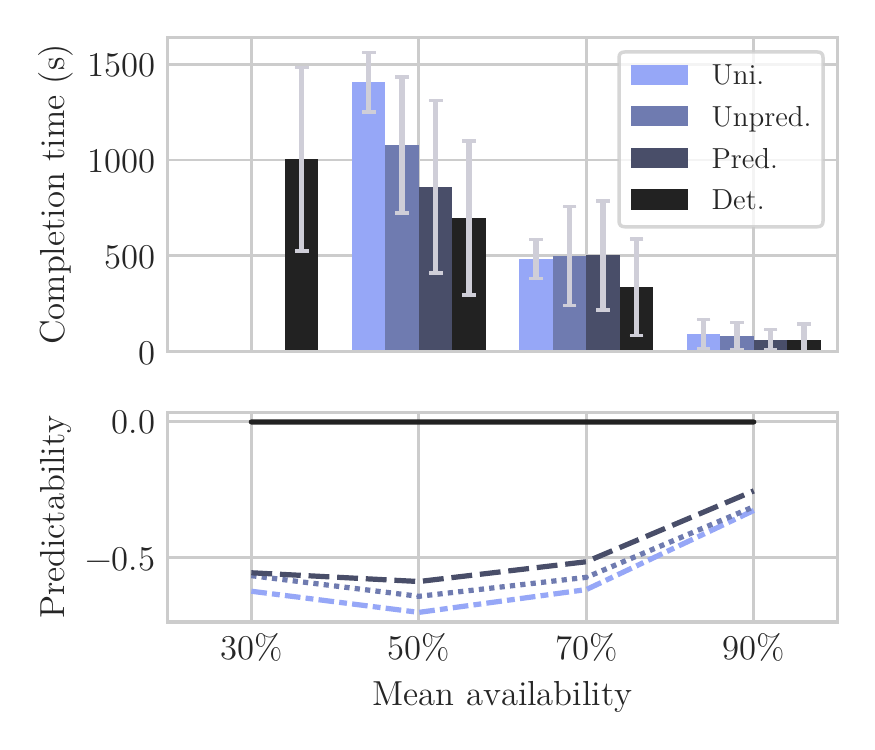}
	\caption{\emph{On top:} File transfer median completion times for each user model, per mean availability $\mu$.
	Error bars represent the interquartile range.
	\emph{On the bottom:} The predictability of each user model per mean availability. Higher is better, as this is a log score.}
	\label{fig:completion_times_f_mean_avail}
\end{figure}

\subsection{Conducted experiments} % (fold)
\label{sub:conducted_experiments}

We evaluated our system on three different regards: its performance under different conditions, its security against colluding attackers, and finally the dynamics of \pors.

\subsubsection{Influence of the users' behaviour} % (fold)
\label{ssub:influence_of_the_user_s_behaviour}

We first study the influence of the user models and mean availability of devices on the file transfer performances.
Towards this goal, we perform an experiment per model described in section~\ref{ssub:users_behavioural_models} and per $\mu \in [0.3, 0.5, 0.7, 0.9]$.
On figure~\ref{fig:completion_times_f_mean_avail}, we display the file transfers median completion times for each of these experiments, along with the predictability of each model.
The top error bars represent the interquartile range (that is the range between the 25th and 75th percentiles of the completion time distribution).
Empty error bars mean that no file exchange succeeded (at $\mu=30\%$, all models fail except the unpredictable (\emph{Unpred.}) one).
Because the file transfer is handled by devices that suffer from churn just as their fellow peers, the transfer times should not be taken literally: they merely serve as a metric to compare outcomes.

We see that the deterministic model stands out in terms of predictability, while the other models follow a similar pattern with their predictability being minimal at $\mu=50\%$.
They are still ordered as was predicted in sec.~\ref{ssub:users_behavioural_models}.
Consequently, the deterministic model always shows better file exchange completion times than the other models.
Most importantly, it is the only model where transfers are entirely completed when devices are only available 30\% of the time.
We also see that the influence of the user model decreases as the network get more available: it is more interesting to perform predictive routing when the risk of dropping messages is big.

\subsubsection{Security measurements} % (fold)
\label{ssub:security_analysis}

% \begin{figure}[t]
% 	\centering
% 	\includegraphics[width=0.8\columnwidth]{figures/attack_on_routes_unpred.pdf}
% 	\caption{Proportion of compromised routes as a function of the percentage of colluding adversaries over the whole network, for both attack types.
% 	We see that as long as the percentage of colluding attackers remains low (which is likely to be the case in our context), only few routes are corrupted.}
% 	% \caption{Success rate of the attack where a colluding adversary owns relays in each hop of an onion route, as a function of the proportion of colluding adversaries over the whole network. On top, statistics are computed with a mean availability of $\mu=70\%$; on the bottom, only the unpredictable (Unpred.) model is displayed.}
% 	\label{fig:attack_all_layers}
% \end{figure}

\begin{figure}[t]
	\centering
	\includegraphics[width=0.8\columnwidth]{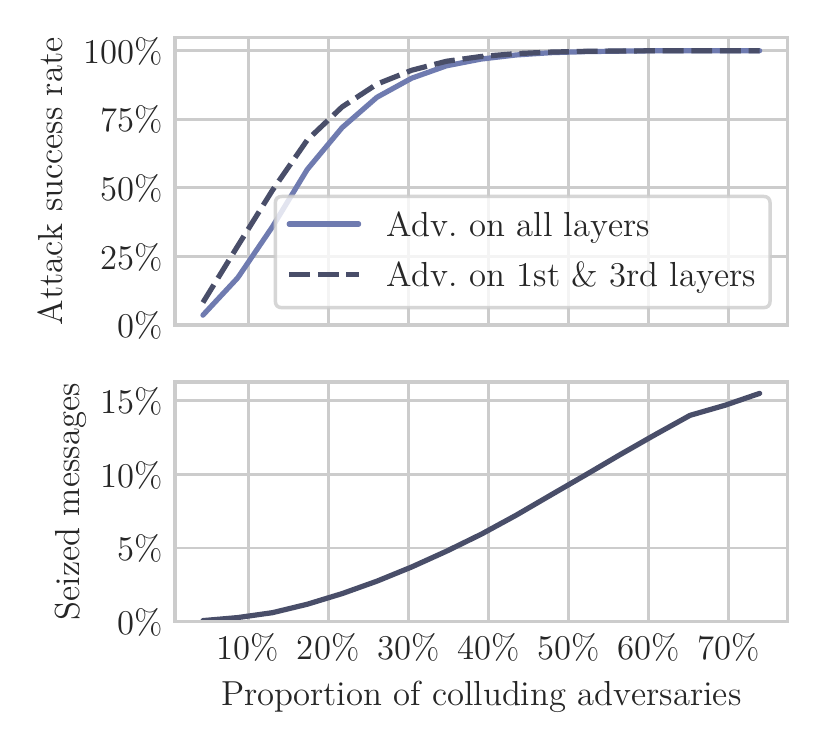}
	\caption{\emph{On top:} Proportion of compromised routes as a function of the percentage of colluding adversaries over the whole network, for both attack types.
	We see that as long as the percentage of colluding attackers remains low (which is likely to be the case in our context), only few routes are corrupted.
	\emph{On the bottom:} Proportion of messages observed by the attacker on each end of the route, considering only routes that were compromised.}
	\label{fig:attack_on_routes}
\end{figure}

% Depending on the proportion of colluding users attacking the system, we notably display on figure~\ref{fig:attack_all_layers} the percentage of routes that were entirely corrupted (cf. sec.~\ref{sub:likelihood_of_having_adversaries_on_each_hop}), and the proportion of messages that were successfully seized during a traffic correlation attack on figure~\ref{fig:attack_traffic_seized_mess} (cf. sec.~\ref{sub:likelihood_of_end_to_end_traffic_correlation}).

Using the logs from the Unpred. model with $\mu=50\%$,
we now study the occurrence of the attacks presented in section~\ref{sec:attack_model}.
Depending on the proportion of colluding users attacking the system, we display on top of figure~\ref{fig:attack_on_routes} the percentage of routes that were corrupted---either on each hop, as discussed in section~\ref{sub:likelihood_of_having_adversaries_on_each_hop}, or only on the first and last hops, as seen in sec.~\ref{sub:likelihood_of_end_to_end_traffic_correlation}.
The bottom of the figure reads the amount of seized messages once routes are compromised, showing how \spores circumvents traffic analysis attacks.
We only showed results for a single experiment/mean availability couple, because the probability of such attacks does not depend on either parameter (the output curves were mostly equal).

All these statistics were computed as follows: knowing that there are $\nusers=25$ users in each experiment, there are always 23 potential adversaries per file exchange (excluding the sender and receiver).
We consider a number $\nadvusers \in [\![1,\cdots,17]\!]$ of evil users, conspiring to de-anonymise the whole network.
For each value of \nadvusers, we computed up to a thousand combinations of conniving users, and counted the number of times they successfully compromised a route, for each route that was created during the experiment.
Note that, doing so, we consider that attackers follow the same churn model as other users.

For the top of figure~\ref{fig:attack_on_routes}, we counted the number of times the set of attackers successfully positioned themselves on a route, normalised by the total number of routes created in the experiment. 
We plotted two curves for the case where an attacker is on the whole route, and when they are only on the route's ends.
%Note that the resulting probability does not measure whether a message was actually received by an attacker on the first layer, as was described in section~\ref{sub:likelihood_of_having_adversaries_on_each_hop}.
The bottom of the figure displays the proportion of messages effectively observed by both ends of a compromised route while attackers perform an end-to-end traffic correlation attack.
%Similarly to the previous figure, we sampled up to 1000 combinations of colluding attackers, for each value of \nadvusers, and took interest in routes that contained adversaries in their first and last layers (the proportion of compromised routes resembles figure~\ref{fig:attack_all_layers} with higher probabilities).
To compute this statistics, for each compromised route, we counted the number of messages that passed through the adversarial relays, normalised by the total number of messages passing through this route.

We firstly see that it suffices to own around 40\% of the network for all created routes to be compromised.
3.6\% of the routes would be entirely compromised by an attacker owning 4.3\% of the nodes.
Although daunting, this observation constitutes a major argument in favour of the multiplication of relays in any onion network.
To resist such de-anonymisation attempts, it us crucial for onion networks to let any participating device partake in the routing.

Where \spores stands out, in terms of security, is on its resilience to traffic analysis attacks: even when an attacker successfully positions themself on both ends of the route,
they can hardly observe 15\% of traffic, even if they subverted most of the network.
Given that most existing end-to-end traffic correlation attacks assume that the eavesdropper sees all the traffic, the multi-path routing severely hampers this attack vector.

% Figure~\ref{fig:attack_all_layers} shows that the amount of routes compromised quickly rises to 100\%, whatever the user model or mean availability, except for $\mu=90\%$.
% Most of the time, if 20\% of the whole network is corrupted by the same attacker, it suffices to de-anonymise more than half of the routes.
% When the mean availability of devices $\mu=90\%$, it gets harder for attackers to compromise routes.
% This can be explicated by the better availability of righteous devices.
% Other than that, figure~\ref{fig:attack_all_layers} mostly motivates the need for a high number of relays in an onion network, so as to ensure that no adversary can subvert a substantial proportion of relays.

% Figure~\ref{fig:attack_traffic_seized_mess}, on the other hand, shows that, in \spores, the traffic correlation attack is not the most daunting issue.
% Even an attacker owning 70\% of the network, positioned on the first and last layers of a route hardly catches one message out of five on both layers.
% When $\nadvusers=3$ (13\% of the network), the adversary observes less than 1\% of the passing messages.
% \commentAL{Hard to conclude without a actual attack scenario. Is seizing 1\% of the messages enough to de-anonymise?}

% \commentAL{Also, why the fuck does $\mu=0.7$ yields more seized messages?}

% subsubsection security_analysis (end)

\subsubsection{Tuning Probabilistic Onion Routes} % (fold)
\label{ssub:tuning_probabilistic_onion_routes}

% \begin{figure}[t]
% 	\centering
% 	\includegraphics[width=0.8\columnwidth]{figures/2_avail_thres_layer_size_end_success_rate_global_newer_descending.pdf}

% 	\commentAL{Modify figure: \textbf{un}availability threshold}
% 	\caption{Boxplots of the layer size per unavailability threshold $\theta$ (left axis), and end-to-end messages exchange success rate per $\theta$ (grey line, right axis).
% 	% We do not take e-squad layers in consideration: as they fail to reach the availability threshold in 33.0\% of the cases.
% 	% The success rate is maximal for $\theta=0.01$: adding more devices per layer adds overhead without benefits, while less devices per layer makes routes unreliable (with a minimum success rate at 9.3\% for $\theta=1$, that is 1 device/layer).
% 	}
% 	\label{fig:devices_per_layer_f_avail_thres}
% \end{figure}

\begin{figure}[t]
	\centering
	\includegraphics[width=0.8\columnwidth]{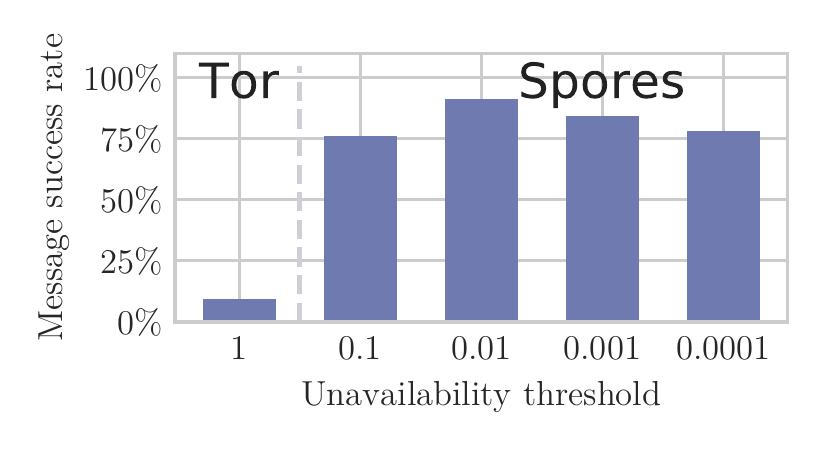}
	\caption{Success rate of the messages transit depending on the unavailability threshold $\theta$. With $\theta=1$, \spores behaves like traditional onion routing with one node per layer.}
	\label{fig:mess_transit}
\end{figure}

We have performed a last experiment using the Unpred. model with $\mu=0.5$, where we varied the unavailability threshold $\theta$ used by the \picklayer route selection function.
We took $\theta=\{1, .1, .01, .001, .0001\}$, and exchanged 50 files, resulting in 10 files per value of $\theta$.
Remember that, when $\theta=1$, the \picklayer function is satisfied as long as the layer's probability of being offline ($\Poff_\layer$) is lower than 1, that is, when there is one device per layer.
At $\theta=1$, the resulting route will resemble traditional onion routing.

To assess the influence of $\theta$, we counted the number of messages that successfully traversed their \por for each parameter value, resulting in the message transit success rates displayed on figure~\ref{fig:mess_transit}.
Given the high churn of the experiment, traditional onion routes ($\theta=1$) only allowed 9.2\% of their traffic to go through. 
We see that adding any number of devices per layer allows to reach 75\% of reliability at least.
The most reliable route occurs when $\theta=.0.1$, where 91\% of the messages go through.
Adding more devices per layer only weakens the route despite the added redundancy: this is explained by the bigger network cost (due to the increased header size) and latency (due to the sequential attempts at reaching the next layer's relays) of adding alternatives.
It is interesting to note that, as $\theta$ shrinks exponentially, the number of relays per layer seems to grow linearly: it is 2.5 for $\theta=0.1$, 4, when $\theta=0.01$, 6 when $\theta=0.001$, and 8 when $\theta=0.0001$.

We see that multi-path routing is a very promising prospect for onion routing over an unreliable network, and that a reasonable threshold is enough to provide maximum route efficiency.
The header size linearly grows (per increments of a symmetric block size) per the number of relays per hop, while the added cryptographic cost of deciphering a handful of envelopes per hop is minimal. 
We hope to improve on our header format in future works to make \pors even more powerful,  \textit{via} the use of Sphinx~\cite{danezis_sphinx_2009} to enable more compact and secure header formats.

\section{Related works} % (fold)
\label{sec:related_works}

Anonymous file sharing between people is not an easy problem, for technical and political reasons~\cite{munroe_file_2011,higgins_troubling_2014}.
In 2014, OnionShare (\url{https://onionshare.org}) solved the issue by proposing a solution over Tor:
one of the two persons (the `server') willing to exchange a file creates an onion service over Tor~\cite{dingledine_tor_2004}, and provides the other person (the `client') with an onion link (a random hash URL finishing with `.onion') pointing to the service.
The client then visits that site through the Tor browser, and can either download from or upload to the server (depending on the configuration mode). Once the transfer is completed, the server tears down the service, leaving no further trace of the file exchange.
OnionShare requires that the server be created \emph{prior} to the file exchange, while \spores requires no such bootstrap, simplifying the exchange. Furthermore, OnionShare leverages Tor, consequently it is inherently susceptible to a variety of traffic analysis attacks~\cite{cai_touching_2012,nasr_compressive_2017,rochet_dropping_2018}.
A variety of proposals attempt to circumvent the attacks by enhancing the route selection process~\cite{sun_counter-raptor_2017,barton_towards_2018,wan_guard_2019}.
Our work, orthogonal to these, takes another approach: we promote a multiplication of relays while being churn tolerant to effectively improve anonymity.
Supporting the same claim, HORNET~\cite{chen_hornet_2015} proposes a new onion routing strategy aiming better performance and resistance to mass surveillance programs by rendering relays stateless.
However, \spores is still more resilient to the aforementioned attacks, though, thanks to our multi-path routing approach.

Some academic proposals, such as Tarzan~\cite{Freedman:2002}, Vuvuzela~\cite{van_den_hooff_vuvuzela_2015} or Loopix~\cite{piotrowska_loopix_2017} do tackle also traffic analysis, and even the Global Passive Adversary (GPA) model where an attacker would listen on all communication pipes.
All of them achieve this feat by generating dummy cover traffic, which we consider undesirable due to the important footprint of such approach, when we target mobile appliances with constrained resources.

We take the most interest in efforts to decentralise anonymity networks, which would allow them to scale and be more resilient.
I2P, being more than a decade years old, has to be cited as a fully P2P anonymity system, comprising 34k daily users~\cite{hoang_empirical_2018}.
Although, the lack of coverage of its security properties does not allow to compare it to other systems.
Recent prospects to allow decentralisation of networks take interest in leveraging blockchain technologies, or trusted computer zones, to realise critical building blocks of decentralised systems.
NextLeap~\cite{halpin_nextleap_2017}, for instance, offers to solve the problem of identifying peers using blockchain technologies.
SGX-Tor~\cite{kim_sgx-tor_2018} proposes to make onion relays more secure by running them in encrypted enclaves; and ConsenSGX suggests that Tor's centralised Directory Authorities consensus could scale to more relay servers using the same components.
\spores does not need trusted computing to warrant its security properties. 
Although, blockchains or trusted enclaves would be an interesting avenue for future developments in user authentication, for instance.

% I2P is a good example of P2P anonymity network. Despite its low media coverage and shallow documentation, it comprises around 34k daily users~\cite{hoang_empirical_2018}.

% Some building blocks such as NextLeap~\cite{halpin_nextleap_2017} would enable taking care of the private-key infrastructure in a decentralised way.
% Trusted computing also could enhance 
% while leveraging on trusted computing enclaves could alleviate the need to share the whole state of a system (e.g. Tor's consensus) without sacrificing security~\cite{sasy_consensgx_2019}.

% The problem is: distributed systems need trust, at least for identity management.
% An interesting avenue for building trust in a fully decentralised network is to leverage the Trusted Execution Environment (TEE) that comes with most recent processors (e.g. Intel's Software Guard eXtensions (SGX) or ARM's TrustZone)~\cite{costan_intel_2016}.
% They are co-processors that work only on encrypted memory, and provide verifiable operation, although at an expensive computing cost.
% Kim et al~\cite{kim_sgx-tor_2018} envisioned to put every Tor relay inside SGX, which would ensure their validity and hide the relays registry inside the TEE's encrypted memory.
% Broader contributions tackle e.g. the problem of encrypting content for a group of peers~\cite{contiu_ibbe-sgx_2018}.

%\commentAL{Potential axes to explore: HORNET/Sphinx about the header format; distributed PKI/sybils; other means to improve Tor's efficiency/security.}

\section{Conclusion} % (fold)
\label{sec:conclusion}

With \spores, we have proposed an anonymous P2P file transfer protocol by revisiting traditional onion routing and leveraging on the people's own devices.
Overall, we have seen that \spores was a sound approach to onion routing in challenging network conditions.
Through its predictive component, it can successfully accomplish file transfers even in the worst connectivity scenarii.
Its security properties are novel, as it is one of the first onion routing approach to finally hinder traffic correlation attacks.
Its design makes it fit for large scale deployments on commodity hardware, which would lower the risk of de-anonymised routes.
Finally, the multi-path routing approach proves its worth when compared to the legacy on an unstable network: the fact of proposing just one alternative node per layer already increases the routes' reliability by 65\%.

In future works, we hope to improve on \spores on several aspects.
Our predictive model could be enhanced; it would firstly require some field studies about people's usage of their e-squads.
We also look upon contributions like Sphinx~\cite{danezis_sphinx_2009} to have more compact and secure header formats.
%Finally, we would like to investigate contributions in distributed PKI to improve the security of such decentralised system.

%The Sphinx mix header format~\cite{danezis_sphinx_2009} could be adapted to enable fixed-size messages despite the several nodes per hop and \spores' statelessness.

% \appendix
% \input{route_creation_appendix.tex}

%-------------------------------------------------------------------------------

\newgeometry{twoside=true, head=13pt,
twocolumn, paperwidth=8.5in, paperheight=11in,
includehead=false, columnsep=2pc,
top=1.1in, bottom=1.1in, inner=0.75in, outer=0.75in,
}

%%bibtex
\bibliographystyle{ACM-Reference-Format}
\bibliography{main}
%%biblatex
%\printbibliography[]

%%%%%%%%%%%%%%%%%%%%%%%%%%%%%%%%%%%%%%%%%%%%%%%%%%%%%%%%%%%%%%%%%%%%%%%%%%%%%%%%
\end{document}